\documentclass{amsart}

\usepackage[utf8]{inputenc}
\usepackage{color,txfonts}
\usepackage[colorlinks=true]{hyperref}

\usepackage{graphicx}
\numberwithin{equation}{section}

\newtheorem{proposition}{Proposition}
\newtheorem{definition}{Definition}

\def \GL{\mathrm{GL}}

\hypersetup{urlcolor=blue, citecolor=red,
   filecolor=magenta,        }

\begin{document}

\title{Klein's programme and Quantum Mechanics}
\author{Jesús  Clemente-Gallardo}
\address{BIFI-Departamento de Física Teórica\\ Edificio I+D-Campus Río
  Ebro\\ Universidad de Zaragoza \\ 
  50018 Zaragoza(SPAIN)\\}
\email{jesus.clementegallardo@bifi.es}

\author{ Giuseppe Marmo}
\address{Dipartimento de Fisica \\
 Università Federico II di Napoli and
  \\
INFN Sezione di Napoli \\
  80126-Napoli (ITALY)}
\email{marmo@na.infn.it}

\maketitle

\begin{abstract}
We review the geometrical formulation of Quantum Mechanics to identify,
according to Klein's programme, the corresponding group of
transformations. For closed systems, it is the unitary group. For open
quantum systems, the semigroup of Kraus maps contains, as a maximal
subgroup, the general linear group. The same group emerges as the
exponentiation of the $C^{*}$--algebra associated with the quantum
system, when thought of as a Lie algebra. Thus, open quantum systems
seem to identify the general linear group as associated with quantum
mechanics and moreover suggest to extend the Klein programme also to
groupoids. The usual unitary group emerges as a maximal compact
subgroup of the general linear group.
\end{abstract}


\section{Introduction: geometric description of Quantum Mechanics}

The Erlangen Programme (see \cite{Klein2008} ) was above all an affirmation
of the key role of groups in geometry.  Klein writes: "In the following third
part of my collected works ... are those papers  involving the concept of
continuous group of transformations".  He announced the main aspects of his
programme as follows: "Given a manifold and a transformation group acting on
it, to investigate those properties of figures on that manifold which are
invariant under all transformations of that group."

In modern language, we could say that each geometry or geometrical structure
is fully characterized by a subgroup of the diffeomorphism group. Different
realizations of the same group give rise to isomorphic geometrical
structures or geometries, i.e., a "geometry" is associated with an abstract
(Lie) group and not with a specific realization.

Therefore, a geometrical description of Quantum Mechanics should identify an
associated group. In this paper, to be able to deal with finite dimensional
Lie groups and corresponding geometrical structures we shall restrict our
considerations to  systems with finite dimensional manifolds of observable and
states. Our presentation should be thought of as providing guidelines to deal
with the  more realistic situation of infinite dimensional states, which
would require a theory of infinite dimensional Lie groups,  not fully
available at present time.

The main point of our paper is that if one takes the $C^{*}$-algebraic approach
to quantum mechanics, the general complex linear group of transformations
emerges as a more fundamental group than its maximal compact subgroup of
unitary transformations. 

To contextualize the search for the group to be associated with Quantum
Mechanics, we shall consider the two main pictures usuallly encountered in
the description of quantum systems, Schrödinger's and Heisenberg's,
from a geometric perspective. We
shall follow the construction which has been developed in the last
years (see for example \cite{Carinena2007b,josegeometry,Clemente-Gallardo2007,ClementeGallardo:2008p614,Ercolessi2010} and references therein).
As a running example we shall consider the case of a two-level system (or a
qubit), and we shall discuss most of the emerging geometrical structures in
this particular situation.

\section{Geometric Quantum Mechanics}
\subsection{The Schrödinger picture}

In the Schrödinger picture, the carrier space $\mathcal{R}(%
\mathcal{H})$ which represents the states of the model is the projective
space corresponding to the Hilbert space $\mathcal{H}$ and, as such, a
Kähler manifold. Therefore the geometrical information of this 
picture is contained in a Riemannian structure, a symplectic structure and a
complex one, which are compatible. Each two of them determines the
third one. This is in agreement with the observation that for linear
invertible transformations, 
\begin{align*}
& \text{symplectic}\cap \text{Riemannian}\simeq \text{unitary} \\
& \text{complex}\cap \text{Riemannian}\simeq \text{unitary} \\
& \text{symplectic}\cap \text{complex}\simeq \text{unitary}
\end{align*}

\subsubsection{The Hilbert space formulation}

The various algebraic structures take the form of tensors defined on the carrier
space. Thus we take the complex vector space of a $n$--level quantum system $%
\mathcal{H}=\mathbb{C}^{n}$ and consider it as a Kähler
manifold: a real vector space $\mathbb{R}^{2n}$ with a complex structure $J$
(a $(1,1)$--tensor field), a Riemannian structure $g$ and a symplectic form $%
\omega $, related as 
\begin{equation}
g(X,Y)=\omega (X,JY),\qquad \forall X,Y\in \mathfrak{X}(M).  \label{eq:19}
\end{equation}
Analogously we can consider the corresponding contravariant tensors $G$ and $
\Lambda $.

If we consider a two level system, the Hilbert space corresponds to $
\mathcal{H}=\mathbb{C}^{2}$ with the usual Hermitian structure 
\begin{equation}
\left\langle z|w\right\rangle =\sum_{k}\bar{z}^{k}w^{k}.  \label{eq:20}
\end{equation}

To consider it as a real vector space we can consider a basis and associated
coordinates $\{z^{1},z^{2}\}$. The real carrier space becomes $\mathbb{R}%
^{4} $ with coordinates $(q^{1},p_{1},q^{2},p_{2})$, which are the real ($%
q^{k}$) and imaginary ($p_{k}$) parts of the complex coordinate $z^{k} $,
for $k=1,2$. In these coordinates, the expression of the tensors defining
the Kähler structure would be as follows:

\begin{itemize}
\item the Riemannian metric corresponds to an Euclidean metric on $\mathbb{R}%
^{4}$: 
\begin{equation}
g=dq^{1}\otimes dq^{1}+dp_{1}\otimes dp_{1}+dq^{2}\otimes
dq^{2}+dp_{2}\otimes dp_{2}  \label{eq:21}
\end{equation}%
or as a bidifferential operator,contravariant (0,2) tensor field: 
\begin{equation}
G=\frac{\partial }{\partial q^{1}}\otimes \frac{\partial }{\partial q^{1}}+%
\frac{\partial }{\partial p_{1}}\otimes \frac{\partial }{\partial p_{1}}+%
\frac{\partial }{\partial q^{2}}\otimes \frac{\partial }{\partial q^{2}}+%
\frac{\partial }{\partial p_{2}}\otimes \frac{\partial }{\partial p_{2}}.
\label{eq:22}
\end{equation}

\item analogously, the symplectic structure can be written as a 2-form 
\begin{equation}
\omega =dq^{1}\wedge dp_{1}+dq^{2}\wedge dp_{2}  \label{eq:23}
\end{equation}%
or as a Poisson bivector field 
\begin{equation}
\Omega =\frac{\partial }{\partial q^{1}}\wedge \frac{\partial }{\partial
p_{1}}+\frac{\partial }{\partial q^{2}}\wedge \frac{\partial }{\partial p_{2}%
}.  \label{eq:24}
\end{equation}

\item finally, the complex structure is written as a (1-1) tensor field 
\begin{equation}
J=dq^{1}\otimes \frac{\partial }{\partial p_{1}}-dp_{1}\otimes \frac{%
\partial }{\partial q^{1}}+dq^{2}\otimes \frac{\partial }{\partial p_{2}}%
-dp_{2}\otimes \frac{\partial }{\partial q^{2}}  \label{eq:25}
\end{equation}
\end{itemize}

As for the operators on $\mathcal{H}$ we can also translate them into
tensorial terms in several ways. One of them is to define the
quadratic functions corresponding to them and write, for any $A\in \mathfrak{%
gl}(\mathcal{H})$ the associated function 
\begin{equation}
f_{A}(\psi )=\frac 12 \langle \psi |A\psi \rangle .
\label{eq:41}
\end{equation}

It is immediate to see that the quadratic functions corresponding to
Hermitian operators are real valued, in the general case they are complex. We shall
denote by $\mathcal{F}_{2}(\mathcal{H})$ and $\mathcal{F}_{2}^{\mathbb{R}}(%
\mathcal{H})$ the set of  quadratic complex valued functions and the
subset of  quadratic real valued functions respectively.

For instance, in our example $\mathcal{H}=\mathbb{C}^{2}\simeq \mathbb{R}^{4}
$ we can consider the set of $2\times 2$ complex matrices which is a
$C^{*}$--algebra, $\mathfrak{gl}(2,\mathbb{C})$  and contains
$\mathrm{GL}(2,\mathbb{C})$ and the real elements correspond to 
Hermitian matrices, which, after multiplication by the imaginary unit,would
correspond to elements in $\mathfrak{u}(2)$. The quadratic functions
corresponding to the Pauli matrices, for instance, would be written as 
\begin{equation}
  \label{eq:80}
  f_{\sigma_{0}}(\psi)=\frac 12 \left
    ((q^{1})^{2}+(q^{2})^{2}+p_{1}^{2}+p_{2}^{2}\right ),
\end{equation}

\begin{equation}
f_{\sigma _{1}}(\psi (q,p))= \frac 12 \langle \psi |\sigma _{1}\psi \rangle
=\left (q^{1}q^{2}+p^{1}p^{2}\right ),  \label{eq:55}
\end{equation}%
\begin{equation}
f_{\sigma _{2}}(\psi (q,p))=\frac 12\langle \psi |\sigma _{2}\psi \rangle
=\left (q^{1}p_{2}-q^{2}p_{1}\right ),  \label{eq:55b}
\end{equation}%
and
\begin{equation}
f_{\sigma _{3}}(\psi )=\frac 12 \langle \psi |\sigma _{3}\psi \rangle
=\frac 12 \left ((q^{1})^{2}-(q^{2})^{2}+p_{1}^{2}-p_{2}^{2}\right )  \label{eq:55c}
\end{equation}%
where $\sigma_{0}=\mathbb{I}_{2}$, the identity matrix in two
dimensions.

On $\mathcal{F}_{2}(\mathcal{H})$ we can export the
algebraic structures the set $\mathfrak{gl}(\mathcal{H})$ is endowed with:

\begin{itemize}
\item the natural Lie algebra structure associated with the commutator $
[A,B]_{\mathfrak{gl}}=(AB-BA) $ is not the most convenient since it does
not define a subalgebra structure on the set of Hermitian operators.
Instead, we can consider  $[A,B] =-i(AB-BA) $ which does define a Lie
algebra structure on $\mathrm{Herm}({\mathcal H})$ and that
 can be realized on ${\mathcal F}_{2}({\mathcal H})$ by using the Poisson
 tensor (\ref{eq:24}):  
\begin{equation}
f_{[A,B]}=\Omega (df_{A},df_{B}):=\{f_{A},f_{B}\};\qquad \forall A,B\in 
\mathrm{Herm}(\mathcal{H}).  \label{eq:42}
\end{equation}%
Last equation is defining a Poisson bracket $\{\cdot ,\cdot \}$ on
$\mathcal{F}_{2}^{\mathbb{R}}$. 

\item The Jordan structure associated to the anti-commutator $%
[A,B]_{+}=AB+BA $ is realized by using the symmetric tensor (\ref{eq:22}): 
\begin{equation}  \label{eq:43}
f_{[A,B]_{+}}=G(df_{A}, df_{B}):=\{ f_{A}, f_{B}\}_{+}; \qquad \forall
A,B\in \mathfrak{gl}(\mathcal{H}).
\end{equation}
Last equation is defining a Jordan bracket $\{ \cdot, \cdot \}_{+}$ on $%
\mathcal{F}_{2}(\mathcal{H})$. This operation is inner on $\mathcal{F}%
_{2}^{\mathbb{R}}$.

\item The associative product of $\mathfrak{gl}(\mathcal{H})$ can be written
by using a combination of both  operations above. Thus as $AB=\frac{1}{2}%
[A,B]_{+}+\frac{i}{2}[A,B]$, we can introduce then a $\star$--product
on the quadratic functions
\begin{equation}
f_{AB}=\frac{1}{2}G(df_{A},df_{B})+\frac{i}{2}\Omega
(df_{A},df_{B}):=f_{A}\star f_{B};\qquad \forall A,B\in \mathrm{Herm}(%
\mathcal{H}).  \label{eq:44}
\end{equation}%
If we extend it by linearity, last equation is defining a new
operation $\star $ on $\mathcal{\ F}_{2}(%
\mathcal{H})$, which is non-local and non-commutative, but it is associative. 
\end{itemize}

With respect to these tensors, we can define two types of vector fields
associated with  quadratic functions (and hence to the elements of $%
\mathfrak{gl}(\mathcal{H})$:

\begin{itemize}
\item Hamiltonian vector fields, associated with quadratic functions
  via  the Poisson structure: 
\begin{equation}  \label{eq:52}
X_{f_{A}}=\Omega(df_{A}, \cdot)=\{ f_{A}, \cdot\}, \qquad f_{A}\in \mathcal{F%
}_{2}(\mathcal{H}).
\end{equation}

\item And gradient vector fields  associated with quadratic functions
  via  the symmetric tensor $G$: 
\begin{equation}  \label{eq:53}
Y_{f_{A}}=G(df_{A}, \cdot)=\{f_{A}, \cdot \}_{+}, \qquad f_{A}\in \mathcal{F}%
_{2}(\mathcal{H}).
\end{equation}
\end{itemize}

Gradient  vector fields associated with the Pauli matrices  via
the quadractic functions on $%
\mathcal{H}=\mathbb{C}^{2}\simeq \mathbb{R}^{4}$ are given by: 
\begin{equation}  \label{eq:56}
Y_{f_{1}}=q^{2}\frac{\partial}{\partial q^{1}}+q^{1}\frac{\partial}{\partial
q^{2}} + p_{2}\frac{\partial}{\partial p_{1}}+ p_{1}\frac{\partial}{\partial
p_{2}}
\end{equation}
\begin{equation}  \label{eq:56b}
Y_{f_{2}}= p_{2}\frac{\partial}{\partial q^{1}}+q^{1}\frac{\partial}{\partial
p_{2}} - p_{1}\frac{\partial}{\partial q^{2}}+ q^{2}\frac{\partial}{\partial
p_{1}}
\end{equation}
\begin{equation}  \label{eq:56c}
Y_{f_{3}}=q^{1}\frac{\partial}{\partial q^{1}}-q^{2}\frac{\partial}{%
\partial q^{2}} + p_{1}\frac{\partial}{\partial p_{1}}- p_{2}\frac{\partial%
}{\partial p_{2}}
\end{equation}
while for the corresponding Hamiltonian vector fieds we find
\begin{equation}  \label{eq:57}
X_{f_{1}}=q^{2}\frac{\partial}{\partial p_{1}}+q^{1}\frac{\partial}{\partial
p_{2}} - p_{2}\frac{\partial}{\partial q^{1}}- p_{1}\frac{\partial}{\partial
q^{2}}
\end{equation}
\begin{equation}  \label{eq:57b}
X_{f_{2}}=p_{2}\frac{\partial}{\partial p_{1}}-q^{1}\frac{\partial}{\partial
q^{2}} - p_{1}\frac{\partial}{\partial p_{2}}- q^{2}\frac{\partial}{\partial
q^{1}}
\end{equation}
\begin{equation}  \label{eq:57c}
X_{f_{3}}=q^{1}\frac{\partial}{\partial p_{1}}-q^{2}\frac{\partial}{%
\partial p_{2}} - p_{1}\frac{\partial}{\partial q^{1}}+ p_{2}\frac{\partial%
}{\partial q^{2}}
\end{equation}
From the compatibility conditions (Eq. (\ref{eq:19}%
)) of the tensors of the Kähler structure, it is
clear  that 
\begin{equation}  \label{eq:54}
Y_{f_{A}}=-J (X_{f_{A}}).
\end{equation}

We notice that the Hamiltonian vector fields close on the Lie algebra of
$\mathfrak{su}(2)$ while the three gradient vector fields transform like a
vector under the action of the Hamiltonian ones. The commutator of two
gradient vector fields is not of gradient type, but it turns  out to
be a Hamiltonian vector field.

\subsubsection{The projective space formulation}

As it is well known the probabilistic interpretation of quantum
Mechanics requires that (pure) states are described by rays and not by
vectors. Therefore the set of pure states is described by the complex
projective space $\mathcal{PH}$. To transfer the
constructed tensorial description to the projective space we must first 
write the relevant objects  in a geometric language. To this aim, we
consider two vector fields

\begin{itemize}
\item the dilation vector field on $\mathcal{H}$, denoted as $\Delta 
$, which in coordinates $(q^{j},p_{k})$ reads 
\begin{equation}
\Delta =\sum_{k}\left( q^{k}\frac{\partial }{\partial q^{k}}+p_{k}\frac{%
\partial }{\partial p_{k}}\right) ,  \label{eq:45}
\end{equation}

\item and the vector field $\Gamma=J(\Delta)$, representing the global
phase change on the Hilbert space $\mathcal{H}$ and which in local
coordinates reads 
\begin{equation}  \label{eq:46}
\Gamma=\sum_{k}\left (q^{k}\frac{\partial}{\partial p_{k}}-p_{k}\frac{%
\partial}{\partial q^{k}}\right )
\end{equation}
\end{itemize}

As they define an integrable distribution we can consider the
corresponding foliation $\mathcal{P}$, the manifold of leaves represents $%
\mathcal{PH}$ in the geometric language. The corresponding fibration $\pi :%
\mathcal{H}_{0}\rightarrow \mathcal{P}$  on the projective space identifies
projectable tensors on $\mathcal{H}_{0}=\mathcal{H}-\{0\}$ \ as those which
are defined on the space of pure states. With this identification we can
avoid using the different charts required to work on $\mathcal{P}$ (as it is
no longer a vector space), and use objects on $\mathcal{H}$, as long as they
are projectable with respect to $\pi $.

Both vector fields have interesting properties. Just to mention some of
them, we can consider their relations with respect to the Kähler structure:

\begin{itemize}
\item The norm of the vector fields $\Gamma$ and $\Delta$  at any
  point  equals the norm of the 
(vector) point at which they are evaluated, i.e. 
\begin{equation}
g(\psi )(\Delta (\psi ),\Delta (\psi ))=\langle \psi |\psi \rangle =g(\psi
)(\Gamma (\psi ),\Gamma (\psi )).  \label{eq:60v}
\end{equation}

\item Analogously, the symplectic area generated by them at a point $\psi
\in {\mathcal H}$ is equal to the norm of the vector (point) 
\begin{equation}
\omega (\Delta ,\Gamma )=\omega (\Delta ,J(\Delta ))=g(\Delta ,\Delta
)=\langle \psi |\psi \rangle .  \label{eq:62}
\end{equation}

\item The two vector fields are orthogonal with respect to $g$: 
\begin{equation}  \label{eq:61}
g(\Delta, \Gamma)=g(\Delta, J(\Delta))=\omega(\Delta,
J^{2}(\Delta))=-\omega(\Delta, \Delta)=0,
\end{equation}
where we used the compatibility condition of the Kähler
structure (Eq. (\ref{eq:19})). This is a general property for any two vector
fields of the form $X$ and $J(X)$, they are always orthogonal to each other.
We shall come back to this point later.
\item They are the gradient and Hamiltonian vector fields associated
  to the quadratc function corresponding the identity matrix,
  i.e., given $f_{0}=\frac12  \sum_{k}(q^{k}) ^{2}+p_{k}^{2})$, we have

\begin{equation}
  \label{eq:81}
  Y_{f_{0}}=\Delta=\sum_{k}\left ( q^{k}\frac{\partial}{\partial
      q^{k}}+ p_{k}\frac{\partial}{\partial p_{k}}\right ),
\end{equation}
and
\begin{equation}
  \label{eq:82}
  X_{f_{0}}=\Gamma=J(\Delta)=\sum_{k}\left
    (q^{k}\frac{\partial}{\partial p_{k}}-p_{k}\frac{%
\partial}{\partial q^{k}}\right ).
\end{equation}
\end{itemize}

In the example of the two level system, the projective space $\mathbb{CP}%
^{1} $ is diffeomorphic to the sphere $S^{2}$, and the corresponding
fibration $\mathbb{C}_{0}^{2}\simeq \mathbb{R}_{0}^{4}\to \mathbb{CP}%
^{1}\simeq S^{2}$ is a generalization of the Hopf fibration $S^{3}\to S^{2}$.

Thus although we can  represent the observables by means of functions, they can
not be those given by Eq. (\ref{eq:41}).  Indeed  they are not
projectable, since they are not homogeneous of degree zero: 
\begin{equation}
\mathcal{L}_{\Delta }f_{A}\neq 0.  \label{eq:47}
\end{equation}%
A possible way out is to use  a
conformal factor: 
\begin{equation}
e_{A}(\psi )=\frac{\langle \psi |A\psi\rangle }{2\langle \psi
|\psi \rangle }  \label{eq:48}
\end{equation}%
With this choice these functions correspond to the physical expectation value of
Hermitian operators, apart from the factor $\frac 12$.

Notice that these functions  contain the spectral information of the
operator to which they are associated. Indeed, it is immediate to verify that

\begin{itemize}
\item the eigenvectors of the operator $A$ correspond to the critical points
of the function $e_{A}$,

\item the values of the function $e_{A}$ at the critical point is
precisely the eigenvalue at the corresponding eigenvector up to a
factor $\frac 12$.
\end{itemize}

For instance, if we consider the function 
\begin{equation}  \label{eq:76}
e_{3}(\psi)=\frac{\langle\psi|\sigma_{3}\psi\rangle}{2\langle
\psi |\psi\rangle},
\end{equation}
its critical points are obtained as those points $(q^{1}, q^{2}, p_{1},
p_{2})\in \mathbb{R}_{0}^{4}$ satisfying 
\begin{equation}  \label{eq:77}
de_{3}=0 \Rightarrow 
\begin{cases}
q^{2}=p_{2}=0\quad q^{1},p_{1}\neq 0 \\ 
q^{1}=p_{1}=0\quad q^{2},p_{2}\neq 0%
\end{cases}%
\end{equation}
The first critical set represents the eigenspace $\mathrm{span}_{\mathbb{C}%
}\left \{ 
\begin{pmatrix}
1 \\ 
0%
\end{pmatrix}
\right \}$ while the last one represents the eigenspace $\mathrm{span}_{%
\mathbb{C}}\left \{ 
\begin{pmatrix}
0 \\ 
1%
\end{pmatrix}
\right \}$.

In what concerns the tensors associated to the Kähler
structure of $\mathcal{H}$ it is evident again that they can not be directly
projected. Indeed, if we consider the contravariant tensors $G$ and
$\Omega $, we notice that they are homogeneous of degree -2:  
\begin{equation}
\mathcal{L}_{\Delta }G=-2G;\qquad \mathcal{L}_{\Delta }\Omega =-2\Omega ;
\label{eq:49}
\end{equation}%
and therefore they can not be projected onto $\mathcal{P}$. Instead, we can
consider alternative degenerate tensor fields as 
\begin{equation}
G_{\mathcal{P}}(\psi )=\langle \psi |\psi \rangle G(\psi )-\left( \Gamma
\otimes \Gamma +\Delta \otimes \Delta \right) (\psi )  \label{eq:50}
\end{equation}%
and 
\begin{equation}
\Omega _{\mathcal{P}}(\psi )=\langle \psi |\psi \rangle \Omega (\psi
)-\left( \Gamma \otimes \Delta -\Delta \otimes \Gamma \right) (\psi ).
\label{eq:51}
\end{equation}%
These tensor fields at $\psi $ are clearly homogeneous of degree zero
and invariant under the action of $\Gamma$ and hence
they are projectable. The factors containing $\Gamma $ and $\Delta $, are
chosen to make them correspond to the tensors of the canonical Kähler
structure of the projective space. From the relations 
seen above it is immediate to verify that one-forms associated to the
vector fields $\Gamma$ and  $\Delta $  by the symplectic or the
Riemannian structures  are in the kernel of $G_{\mathcal{P}}$
and $\Omega _{\mathcal{P}}$. Indeed, consider the mappings $\hat{\omega}:%
\mathfrak{X}(\mathcal{H})\rightarrow \Lambda ^{1}(\mathcal{H})$ 
\begin{equation}
\hat{\omega}(X):Y\mapsto \omega (X,Y),\qquad \forall Y\in \mathfrak{X}(%
\mathcal{H}),  \label{eq:64}
\end{equation}%
and analogously $\hat{g}:\mathfrak{X}(M)\rightarrow \Lambda ^{1}(\mathcal{H})
$ 
\begin{equation}
\hat{g}(X):Y\mapsto g(X,Y),\qquad \forall Y\in \mathfrak{X}(\mathcal{H}).
\label{eq:65}
\end{equation}%
Associated with $\Delta $ and $\Gamma $ we have their Riemannian or
symplectic dual forms, which can be seen to be related via $-J$: 
\begin{equation}
\hat{\omega}(\Delta ):Y\mapsto \omega (\Delta ,Y)=-\omega (\Delta
,J(J(Y)))=-g(\Delta ,J(Y))=-\hat{g}(\Delta )\circ J(Y),\quad \forall Y\in 
\mathfrak{X}(\mathcal{H})  \label{eq:66}
\end{equation}%
and analogously 
\begin{equation}
\hat{\omega}(\Gamma ):Y\mapsto \omega (\Gamma ,Y)=\omega (\Gamma ,J(J(Y)))=-%
\hat{g}(\Gamma )\circ J(Y),\quad \forall Y\in \mathfrak{X}(\mathcal{H}).
\label{eq:67}
\end{equation}
The way  $\Gamma$ and $\Delta$ appear  in $G_{\mathcal{P}}$ and $%
\Omega_{\mathcal{P}}$ ensures that $\hat g(\Delta)$ and $\hat g(\Gamma)$ are
in their kernels. 

With respect to these tensors, we can also characterize   functions
$e_{A}$ (Eq. (\ref{eq:48})) as the functions on  $\mathcal{H}_{0}$
which are the pullback of functions on $%
\mathcal{PH}$ such that their associated Hamiltonian fields are also
Killing. We shall denote the set of these functions as $\mathcal{E}(%
\mathcal{H})$.

We can also study the gradient and Hamiltonian vector fields of the set of
functions $\mathcal{E}(\mathcal{H})$ with respect to these projectable
tensors. We should consider then vector fields as 
\begin{equation}  \label{eq:57v}
\mathcal{Y}_{A}=G_{\mathcal{P}}(de_{A}, \cdot); \qquad e_{A}\in \mathcal{E}(%
\mathcal{H}),
\end{equation}
for gradient vector fields and 
\begin{equation}  \label{eq:58}
\mathcal{X}_{A}=\Omega_{\mathcal{P}}(de_{A}, \cdot); \qquad e_{A}\in 
\mathcal{E}(\mathcal{H})
\end{equation}
for Hamiltonian ones.

For the example of the two level system we obtain: 
\begin{align}  \label{eq:59}
\mathcal{Y}_{e_{1}}=&\left (q^{2}-\frac{2q^{1}(p_{1}p_{2}+q^{1}q^{2})}{%
(q^{1})^{2}+(q^{2})^{2}+p_{1}^{2}+p_{2}^{2}}\right ) \frac{\partial }{%
\partial q^{1}}+ \left (q^{1}-\frac{2q^{2}(p_{1}p_{2}+q^{1}q^{2})}{%
(q^{1})^{2}+(q^{2})^{2}+p_{1}^{2}+p_{2}^{2}}\right )\frac{\partial }{%
\partial q^{2}}+  \notag \\
& \left (p_{2}-\frac{2p_{1}(p_{1}p_{2}+q^{1}q^{2})}{%
(q^{1})^{2}+(q^{2})^{2}+p_{1}^{2}+p_{2}^{2}}\right ) \frac{\partial }{%
\partial p_{1}}+ \left (p_{1}-\frac{2p_{2}(p_{1}p_{2}+q^{1}q^{2})}{%
(q^{1})^{2}+(q^{2})^{2}+p_{1}^{2}+p_{2}^{2}}\right ) \frac{\partial }{%
\partial p_{2}} \\
=& Y_{f_{1}}-2e_{1}\Delta;
\end{align}

\begin{align}  \label{eq:59b}
\mathcal{Y}_{e_{2}}=&\left (p_{2}+\frac{2q^{1}(-p_{2}q^{1} +p_{1}q^{2})}{%
(q^{1})^{2}+(q^{2})^{2}+p_{1}^{2}+p_{2}^{2}}\right ) \frac{\partial }{%
\partial q^{1}}+ \left (-p_{1}+\frac{2q^{2}(-p_{2}q^{1} +p_{1}q^{2})}{%
(q^{1})^{2}+(q^{2})^{2}+p_{1}^{2}+p_{2}^{2}}\right ) \frac{\partial }{%
\partial q^{2}}+  \notag \\
& \left (-q^{2}+\frac{2p_{1}(-p_{2}q^{1} +p_{1}q^{2})}{%
(q^{1})^{2}+(q^{2})^{2}+p_{1}^{2}+p_{2}^{2}}\right ) \frac{\partial }{%
\partial p_{1}}+ \left (q^{1}-\frac{2p_{2}(-p_{2}q^{1} +p_{1}q^{2})}{%
(q^{1})^{2}+(q^{2})^{2}+p_{1}^{2}+p_{2}^{2}}\right ) \frac{\partial }{%
\partial p_{2}} \\
=& Y_{f_{2}}-2e_{2}\Delta;
\end{align}
\begin{align}  \label{eq:59c}
\mathcal{Y}_{e_{3}}=&\left ( \frac{2q^{1}(p_{2}^{2}+(q^{2})^{2})}{%
(q^{1})^{2}+(q^{2})^{2}+p_{1}^{2}+p_{2}^{2}}\right )\frac{\partial }{%
\partial q^{1}}+ \frac 12\left ( \frac{-4q^{2}(p_{1}^{2}+(q^{1})^{2})}{%
(q^{1})^{2}+(q^{2})^{2}+p_{1}^{2}+p_{2}^{2}}\right ) \frac{\partial }{%
\partial q^{2}}+  \notag \\
&\left ( \frac{2p_{1}(p_{2}^{2}+(q^{2})^{2})}{%
(q^{1})^{2}+(q^{2})^{2}+p_{1}^{2}+p_{2}^{2}}\right )\frac{\partial }{%
\partial p_{1}}+ \left ( \frac{-2p_{2}(p_{1}^{2}+(q^{1})^{2})}{%
(q^{1})^{2}+(q^{2})^{2}+p_{1}^{2}+p_{2}^{2}}\right )\frac{\partial }{%
\partial p_{2}} \\
=& Y_{f_{3}}-2e_{3}\Delta
\end{align}
for the gradient vector fields and

\begin{align}  \label{eq:60}
\mathcal{X}_{e_{1}}=&\left (-p_{2}+\frac{2p_{1}(p_{1}p_{2}+q^{1}q^{2})}{%
(q^{1})^{2}+(q^{2})^{2}+p_{1}^{2}+p_{2}^{2}}\right ) \frac{\partial }{%
\partial q^{1}}+ \left (-p_{1}+\frac{2p_{2}(p_{1}p_{2}+q^{1}q^{2})}{%
(q^{1})^{2}+(q^{2})^{2}+p_{1}^{2}+p_{2}^{2}}\right ) \frac{\partial }{%
\partial q^{2}}+  \notag \\
& \left (q^{2}-\frac{2q^{1}(p_{1}p_{2}+q^{1}q^{2})}{%
(q^{1})^{2}+(q^{2})^{2}+p_{1}^{2}+p_{2}^{2}}\right ) \frac{\partial }{%
\partial p_{1}}+ \left (q^{1}-\frac{2q^{2}(p_{1}p_{2}+q^{1}q^{2})}{%
(q^{1})^{2}+(q^{2})^{2}+p_{1}^{2}+p_{2}^{2}}\right ) \frac{\partial }{%
\partial p_{2}}\\
=& X_{f_{1}}-2e_{1}\Gamma;
\end{align}

\begin{align}  \label{eq:60b}
\mathcal{X}_{e_{2}}=&\left (q^{2}-\frac{2p_{1}(-p_{2}q^{1} +p_{1}q^{2})}{%
(q^{1})^{2}+(q^{2})^{2}+p_{1}^{2}+p_{2}^{2}}\right ) \frac{\partial }{%
\partial q^{1}}+ \left (-q^{1}+\frac{2p_{2}(-p_{2}q^{1} +p_{1}q^{2})}{%
(q^{1})^{2}+(q^{2})^{2}+p_{1}^{2}+p_{2}^{2}}\right ) \frac{\partial }{%
\partial q^{2}}+  \notag \\
& \left (p_{2}-\frac{2q^{1}(-p_{2}q^{1} +p_{1}q^{2})}{%
(q^{1})^{2}+(q^{2})^{2}+p_{1}^{2}+p_{2}^{2}}\right )\frac{\partial }{%
\partial p_{1}}+ \left (-p_{1}+\frac{%
2q^{2}(-p_{2}q^{1} +p_{1}q^{2})}{%
(q^{1})^{2}+(q^{2})^{2}+p_{1}^{2}+p_{2}^{2}}\right ) \frac{\partial }{%
\partial p_{2}}\\
=& X_{f_{2}}-2e_{2}\Gamma;
\end{align}
\begin{align}  \label{eq:60c}
\mathcal{X}_{e_{3}}=&\left ( \frac{-2p_{1}(p_{2}^{2}+(q^{2})^{2})}{%
(q^{1})^{2}+(q^{2})^{2}+p_{1}^{2}+p_{2}^{2}}\right )\frac{\partial }{%
\partial q^{1}}+ \left ( \frac{2p_{2}(p_{1}^{2}+(q^{1})^{2})}{%
(q^{1})^{2}+(q^{2})^{2}+p_{1}^{2}+p_{2}^{2}}\right )\frac{\partial }{%
\partial q^{2}} +  \notag \\
&\left ( \frac{2q^{1}(p_{2}^{2}+(q^{2})^{2})}{%
(q^{1})^{2}+(q^{2})^{2}+p_{1}^{2}+p_{2}^{2}}\right )\frac{\partial }{%
\partial p_{1}}+ \left ( \frac{-2q^{2}(p_{1}^{2}+(q^{1})^{2})}{%
(q^{1})^{2}+(q^{2})^{2}+p_{1}^{2}+p_{2}^{2}}\right )\frac{\partial }{%
\partial p_{2}}\\
=& X_{f_{3}}-2e_{3}\Gamma
\end{align}
for the Hamiltonian ones.

These vector fields are tangent to the three dimensional sphere
$S^{3}$  of normalized vectors.

A remarkable property of these vector fields is that either one  of the two
families $\{\mathcal{X}_{f_{1}},\mathcal{X}_{f_{2}},\mathcal{X}_{f_{3}}\}$
or $\{\mathcal{Y}_{f_{1}},\mathcal{Y}_{f_{2}},\mathcal{Y}_{f_{3}}\}$, after
projection, span the tangent bundle of the sphere $S^{2}$. Besides, as we
saw above, they are pairwise orthogonal to each other, i.e. 
\begin{equation}
\mathcal{Y}_{e_{k}}\perp \mathcal{X}_{e_{k}},\qquad k=1,2,3.  \label{eq:63}
\end{equation}

Another  interesting property which we will use in the following is the fact
that the Lie algebra generated by the union of both families, i.e.
the Lie algebra generated by
$\{\mathcal{X}_{f_{1}},\mathcal{X}_{f_{2}},\mathcal{X}_{f_{3}},
\mathcal{Y}_{f_{1}},\mathcal{Y}_{f_{2}},\mathcal{Y}_{f_{3}}\}$ is
isomorphic to the special Lie algebra $\mathfrak{sl}(2, \mathbb{C})$
(also isomorphic to the Lie algebra of the Lorentz group).
If we consider the algebra generated by the vector fields defined on
${\mathcal H}$, we see that $\{ X_{f_{1}}, X_{f_{2}}, X_{f_{3}}, Y_{f_{1}},
Y_{f_{2}}, Y_{f_{3}}\}$ generate again the Lie algebra
$\mathfrak{sl}(2, \mathbb{C})$. If we include the vector fields
$Y_{f_{0}}\propto \Delta$ and $X_{f_{0}}\propto \Gamma$, the set 
 generates the full $\mathfrak{gl}(n, \mathbb{C})$.

\begin{figure}[h]
\centering
\includegraphics[width=10cm]{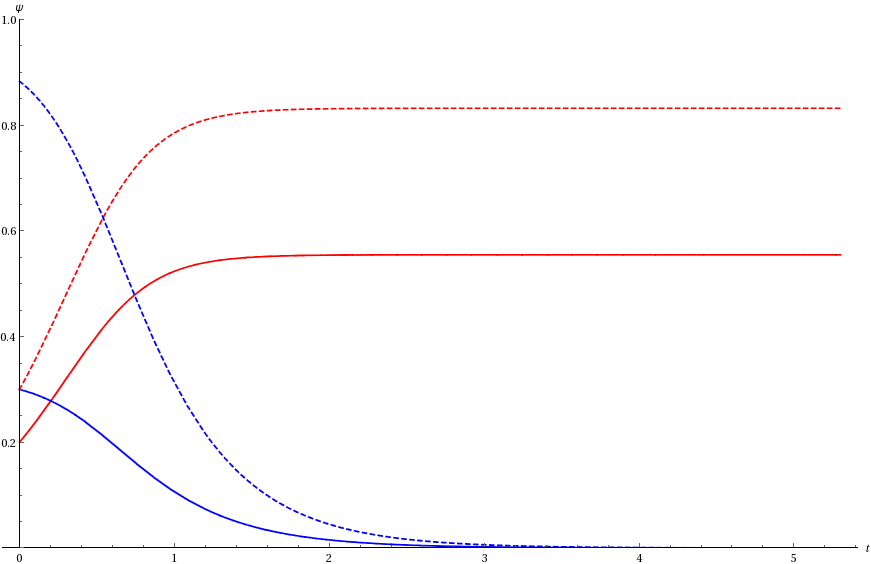}
\caption{Flow of the gradient vector field $\mathcal{Y}_{e_{3}}$ from the
point on the unit sphere with coordinates $q^{1}=0.2$, $q^{2}=0.3$, $%
p_{1}=0.3$ . The red solid line is the flow
for $q^{1}$, the dashed red line $p_{1}$, the blue solid line $q^{2}$ and
the blue dashed line is $p_{2}$. }
\label{fig:1}
\end{figure}

\begin{figure}[h!]
\centering
\includegraphics[width=11cm]{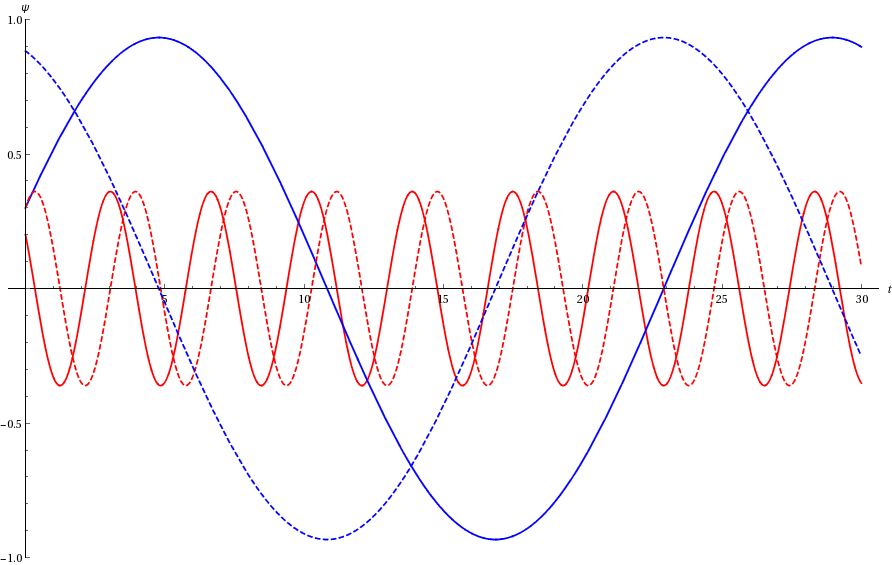}
\caption{Flow of the Hamiltonian vector field $\mathcal{X}_{e_{3}}$ from the
point on the unit sphere with coordinates $q^1=0.2$, $q^2=0.3$,
$p_1=0.3$.  The red solid line is the flow for $q^{1}$,
the dashed red line $p_{1}$, the blue solid line $q^{2}$ and the blue dashed
line is $p_{2}$. }
\label{fig:2}
\end{figure}

As a curiosity we can study the flow associated to these fields. For
instance, if we consider vector fields $\mathcal{X}_{e_{3}}$ and $\mathcal{\
Y}_{e_{3}}$ and represent their flows from an initial condition $\psi _{0}$
we obtain the flows presented in Figures \ref{fig:1} and \ref{fig:2}.

\begin{figure}[h!]
\centering
\includegraphics[width=11cm]{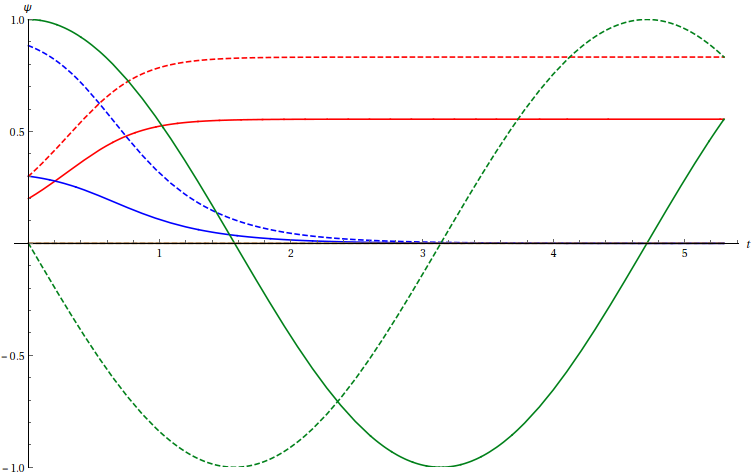}
\caption{Flow of the gradient vector field $\mathcal{Y}_{e_{3}}$ from the
point on the unit sphere with coordinates $q^1=0.2$, $q^2=0.3$ and
$p_1=0.3$ and flow of $\Gamma$ (in green)  from the point
$(1,0,0,0)$. We can see how both flows coincide at one point, thus
proving that the limit point of the gradient flow is in the same
complex ray as  $(1,0,0,0)$. }
\label{fig:3b}
\end{figure}

In Figure \ref{fig:1} we can see how the flow takes us towards one of the
eigenspaces of the operator $\sigma_{3}$, in particular the one generated by 
$\psi= 
\begin{pmatrix}
1 \\ 
0%
\end{pmatrix}
$. In this particular numerical example the limit point is $q^{1}=0.5547$
and $p_{1}=0.83205$, with the remaining two coordinates vanishing, i.e., the
system is selecting the eigenvector $\psi= 
\begin{pmatrix}
0.5547+i\, 0.83205 \\ 
0%
\end{pmatrix}%
$.
But this point is obviously connected by the flow of $\Gamma$ with the point
$(1,0,0,0)$, as it can be seen in Figure \ref{fig:3b}. This is just
reflecting that both points are on an integral curve of $\Gamma$.

\begin{figure}[h]
\centering
\includegraphics[width=11cm]{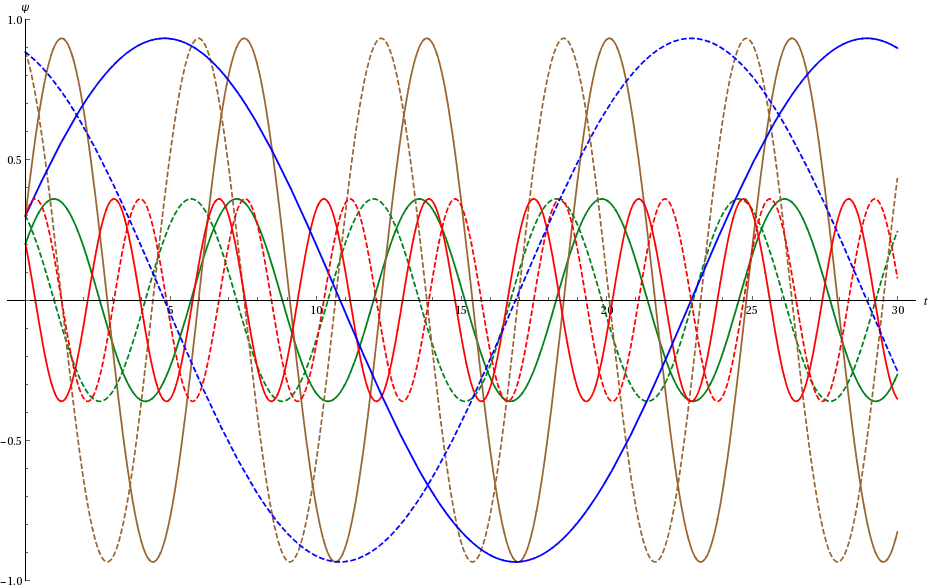}
\caption{Flow of the Hamiltonian vector field $\mathcal{X}_{e_{3}}$ and the
vector field $\Gamma $ from the point on the unit sphere with coordinates $%
q^{1}=0.2$, $q^{2}=0.3$, $p_{1}=0.3$. The
red and blue lines represent the flow of $\mathcal{X}_{e_{3}}$, while the
green and brown lines represent the flow of $\Gamma $. It is simple to
verify that there are no common points, i.e. solid red and solid green lines
do not intersect at the same time that dashed red and green, solid blue and
brown and dashed blue and brown.}
\label{fig:3}
\end{figure}

The flow of the Hamiltonian vector field depicted in Fig \ref{fig:2} is
completely different. We see how the evolution of all four coordinates is
periodic, with two different frequencies, one for $q^{1},p_{1}$ and the
other for $q^{2},p_{2}$.  We can also
verify in Figure \ref{fig:3} that the flow is projecting onto the projective
space by checking that the flow commutes with the flows of the vector fields $%
\Gamma $ and $\Delta$. 

\subsubsection{Alternative structures and representations of the unitary
group}

We notice that the relevant tensor fields associated with the geometry of
the projective space are invariant under  the unitary group. An important
comment is however in order. The specific form of the Hermitian structure (%
\ref{eq:20}) which selects the realization of the unitary group acting on the Hilbert space in
the defining representation is not the only possible one. Eq. \ref{eq:20}
represents the Hermitian structure of the abstract Hilbert space once we
have chosen a basis which defines a one-to-one correspondence between the Hilbert
space ${\mathcal{H}}$ and $\mathbb{C}^{2}$. But we could also consider a
different realization of the Hermitian structure, say in the same basis.
Consider, for instance, a scalar product such as: 
\begin{equation}
\left\langle 
\begin{pmatrix}
z_{1} \\ 
z_{2}%
\end{pmatrix}%
,%
\begin{pmatrix}
w_{1} \\ 
w_{2}%
\end{pmatrix}%
\right\rangle _{\alpha }=\alpha \bar{z}_{1}w_{1}+(2-\alpha )\bar{z}%
_{2}w_{2};\qquad 0<\alpha \leq 1  \label{eq:26}
\end{equation}

This operation defines clearly a complex scalar product on $\mathbb{C}^{2}$
and defines an alternative Hilbert space structure. With it we obtain a
different realization of the unitary group as the isometry group of the new
structure, this new realization possesses all the \textquotedblleft
abstract\textquotedblright\ properties which characterize the unitary group.
At the level of the Lie algebra realization, we would have the vector fields
associated with those matrices which are Hermitian with respect to the
matrix representation 
\begin{equation}
K=%
\begin{pmatrix}
\alpha  & 0 \\ 
0 & 1-\alpha 
\end{pmatrix}%
,  \label{eq:27}
\end{equation}%
i.e., those matrices $A$ satisfying 
\begin{equation}
A^{\dagger }K A=K .
\end{equation}

The point we would like to emphasize is that the "geometry" which a given
group identifies is associated with the "abstract group" and not with a
specific realization.

\subsection{The Heisenberg picture}

\subsubsection{The algebraic formulation}

In a modern language, Heisenberg picture uses as the relevant carrier space
to describe a quantum system  a $C^{\ast }$--algebra $\mathcal{A}$.

\begin{definition}
 A $C^{*}$--algebra is a Banach algebra
over $\mathbb{C}$ with a map  $*: {\mathcal A}\to {\mathcal A}$ satisfying
\begin{itemize}
\item $*$ is an involution, i.e.,  ($(a^{*})^{*}$ for all
  $a\in {\mathcal A}$)
\item the effect of the involution on the algebra structure is given
  as
  \begin{itemize}
  \item $(a+b)^{*}=a^{*}+b^{*}$ for any $a,b\in {\mathcal A}$,
  \item $(ab)^*=b^{*}a^{*}$ for any $a,b\in {\mathcal A}$,
  \item $(\lambda a)^{*}=\bar \lambda a^{*}$ for any
    $\lambda\in\mathbb{C}$ ($\bar \lambda$ represents the complex
    conjugate) and any $a\in {\mathcal A}$,
  \item $\| a a^{*}\|=\|a\|\|a^{*}\|$ for any $a\in {\mathcal A}$.
  \end{itemize}
\end{itemize}
 
\end{definition}

On this carrier space, states correspond to positive and  normalized linear functionals $\rho $ and
observables to the real subalgebra corresponding to the elements $A\in 
\mathcal{A}$ which are stable under the $\ast $--operation. These correspond
to the real elements of the algebra.

In the case of the two level system which we studied above, the $C^{\ast }$
algebra $\mathcal{A}$ corresponds to the $2\times 2$ complex matrices, i.e. $%
\mathcal{A}=M_{2}(\mathbb{C})$, with respect to the usual associative
product. The $\ast $ operation corresponds to the adjoint $A\mapsto
A^{\dagger }$ and the real elements correspond to the self-adjoint matrices $%
A^{\dagger }=A$. This set, $\mathrm{Herm}({\mathcal H})$, is then isomorphic to the Lie algebra of the
unitary group $\mathfrak{u}(2)\subset M_{2}(\mathbb{C})\simeq \mathfrak{gl}(2,
\mathbb{C})$, the isomorphism being just a multiplication by the
imaginary unit (since the Lie algebra $\mathfrak{u}(2)$ is constituted by
anti-Hermitian operators). 
 The norm required to make $\mathcal{A}$ into a Banach
algebra can be given by the trace $\Vert A\Vert ^{2}=\mathrm{Tr}(A^{\dagger
}A)$ which comes from the  scalar product on the Lie algebra $\mathfrak{gl}
({\mathcal H})$ $\langle A|B\rangle=\mathrm{Tr}(A^{\dagger}B)$ and hence
on  the unitary algebra. It is immediate to verify 
that these objects satisfy the conditions above.

To make comparison easier with the Schrödinger picture, we can
consider a description in terms of a Lie-Jordan(-Banach) algebra $\mathcal{L}
$ whose complexification $\mathcal{\ L}^{\mathbb{C}}$ is isomorphic to $%
\mathcal{A}$. We define thus

\begin{definition}
A vector space ${\mathcal L}$ endowed with a Jordan algebra structure $\circ$ and a
Lie structure $[
  \cdot, \cdot ]$, such that  $\forall a, b, c\in \mathcal{L}$: 
  \begin{itemize} 
  \item the Lie-adjoint define derivations of the Jordan operation,
    i.e,  $[a, b\circ c]=[a,b]\circ c +b\circ [a, c]$
  \item the associators of the two operations are proportional to each
    other $(a\circ b)\circ c-a\circ (b\circ c)=\hbar^{2}\left (
      [[a,b],c]-[a,[b,c]] \right )$ where
    $\hbar\in \mathbb{R}$,
  \end{itemize}
is called  a \textbf{Lie-Jordan algebra}.
\end{definition}

\begin{definition}
A Lie-Jordan algebra $\mathcal{L}$ endowed with a norm $\| \cdot
\|$ such that ${\mathcal L}$ is complete and satisfies
\begin{itemize}
\item $\| a\circ b\|\leq \|a\|\|b\|$
\item $\| [a,b]\|\leq |\hbar |^{{-1}}\|a\|\|b\|$
\item $\| a^{2}\|=\|a\|^{2}$
\item $\| a^{2}\|\leq \|a^{2}+b^{2}\|$
\end{itemize}
for any $a, b\in \mathcal{L}$, is called a \textbf{Lie-Jordan-Banach (LJB) algebra}.
  \end{definition}

In the case of a $n$--level quantum system, the LJB algebra $\mathcal{L}$
becomes the set of Hermitian operators $\mathrm{Herm}({\mathcal H})$,
which is isomorphic to the unitary algebra $\mathfrak{u}(n)$, which can be
also identified with its dual $\mathfrak{u}^{*}(n)$. From the practical point of
view, we shall consider all our objects defined on this vector space. The two
products $[\cdot, \cdot ]$ and $\circ$ arise as the skew-symmetric and
symmetric part (respectively) of the associative product defining the $C^{*}$%
--algebra structure of $\mathcal{\ A}$.

\subsubsection{Geometrical formalism and the momentum map}

One important convention we shall use in the following is the
identification of the dual space $\mathfrak{u}^{*}({\mathcal H})$ with the
set of Hermitian operators. Thus, given any
$\xi^{\dagger}=\xi$ and $ T^{\dagger}=-T$,  
\begin{equation}
  \label{eq:75}
  \langle \xi | T\rangle=\frac i2\mathrm{Tr}(\xi T).
\end{equation}
Notice that under this identification $\mathfrak{u}^{*}({\mathcal H})$
becomes a Lie algebra with the bracket
\begin{equation}
  \label{eq:68}
  [\xi_{1}, \xi_{2}]:=-i(\xi_{1}\xi_{2}-\xi_{2}\xi_{1}), \qquad
  \forall \xi_{1}, \xi_{2}\in \mathfrak{u}^{*}({\mathcal H}),
\end{equation}
which is isomorphic to $\mathfrak{u}({\mathcal H})$, by the isomorphism
\begin{equation}
  \label{eq:83}
  \mathfrak{u}^{*}({\mathcal H})\ni \xi\mapsto \hat \xi= -i\xi \in \mathfrak{u}({\mathcal H}).
\end{equation}
 Analogously, we can
also export the scalar product from  $\mathfrak{u}({\mathcal H})$ to
$\mathfrak{u}^{*}({\mathcal H})$ and define
\begin{equation}
  \label{eq:78}
  \langle \xi_{1},
  \xi_{2}\rangle_{\mathfrak{u}^{*}}=\frac 12 \mathrm{Tr}(\xi_{1}\xi_{2})\qquad
  \forall \xi_{1}, \xi_{2}\in \mathfrak{u}^{*}({\mathcal H}). 
\end{equation}
This scalar product allows us to identify linear functionals on
$\mathfrak{u}^{*}({\mathcal H})$ as elements on the Lie algebra
$\mathfrak{u}({\mathcal H})$, recovering the isomorphism (\ref{eq:83}):
\begin{equation}
  \label{eq:100}
  \mathfrak{u}^{*}({\mathcal H})\ni  A \to \hat A :=\langle A, \cdot
  \rangle_{\mathfrak{u}^{*}}=-iA\in \mathfrak{u}({\mathcal H})
\end{equation}

From the tensorial point of view, it is possible to encode the
symmetric and skew-symmetric products of  Hermitian matrices  in two
tensors $R$ and $\Lambda$ defined on the dual space
$\mathfrak{u}^{*}(n)$ as  
\begin{equation}  \label{eq:29}
R(\xi)(d\hat A, d\hat B)=\langle \xi | A\circ B \rangle_{\mathfrak{u}^{*}}, \quad \forall \xi
\in \mathfrak{u}^{*}({\mathcal H}), A, B \in \mathrm{Herm}({\mathcal H})
\end{equation}
and 
\begin{equation}  \label{eq:29b}
\Lambda(\xi)(d\hat A, d\hat B)=\langle \xi | [A, B] \rangle_{\mathfrak{u}^{*}}, \quad \forall
\xi \in \mathfrak{u}^{*}({\mathcal H}), A, B \in \mathrm{Herm}({\mathcal H}).
\end{equation}

Tensor $R$ encodes the Jordan structure while $\Lambda $ corresponds to the
Lie structure on the dual of the unitary algebra. The dynamics,
corresponding to Heisenberg's equation, arises as the integral curve of the
vector field $X_{\hat{H}}\in \mathfrak{X}(\mathfrak{u}^{\ast }({\mathcal H}))$ which is
the Hamiltonian vector field corresponding to the operator $H$: 
\begin{equation}
X_{\hat{H}}=\Lambda (d\hat{H},\cdot ).  \label{eq:72}
\end{equation}
Integral curves are the orbits of a one parameter subgroup
in the co-adjoint action of the unitary group on $\mathfrak{u}^{\ast }({\mathcal H})$.
This property shall be used in the next Section.

It is possible to relate these tensors with those defined on
${\mathcal H}$ and ${\mathcal P}$ which encode the Schrödinger formalism. The
first relevant aspect is the defining action of the unitary group on
the Hilbert space ${\mathcal H}$ and the corresponding projective space
${\mathcal P}$:
\begin{equation}
  \label{eq:83b}
  \Phi:U({\mathcal H})\times {\mathcal H}\to {\mathcal H}; \qquad 
\Phi_{{\mathcal P}}:U({\mathcal H})\times {\mathcal P}\to {\mathcal P}
\end{equation}

By definition, $\Phi$ acts as isometries of the Hermitian structure of
${\mathcal H}$, and therefore, preserves the Kähler structure
$(g,\omega,J)$. Hence the action $\Phi$ is symplectic and must admit a
momentum map. The same happens with $\Phi_{{\mathcal P}}$. Thus we have
two projections
\begin{equation}
  \label{eq:84}
  \mu:{\mathcal H}\to \mathfrak{u}^{*}({\mathcal H}); \qquad \psi\mapsto
  P_{\psi}=\frac 12|\psi\rangle\langle\psi|
\end{equation}
and
\begin{equation}
  \label{eq:85}
  \mu_{{\mathcal P}}:{\mathcal P}\to \mathfrak{u}^{*}({\mathcal H}); \qquad
  [\psi]\mapsto \rho_{\psi}=\frac{|\psi\rangle\langle\psi|}{2\langle\psi|\psi\rangle}
\end{equation}


The fundamental
vector fields of these actions correspond to the Hamiltonian vector
fields $X_{f_{A}}$ on ${\mathcal H}$ and ${\mathcal X}_{e_{A}}$ on ${\mathcal P}$,
for any $A\in \mathrm{Herm}({\mathcal H})$. 
We know that we can associate linear functions  on $\mathfrak{u}^{*}({\mathcal H})$ to operators in
$\mathrm{Herm}({\mathcal H})$  (or equivalently on $\mathfrak{u}({\mathcal
  H})$) as
\begin{equation}
  \label{eq:86}
  \mathrm{Herm}({\mathcal H})\ni A\mapsto \langle 
 P_{\psi} |A \rangle_{\mathfrak{u}^{*}}=F_{A}(P_{\psi})
\end{equation}
and
\begin{equation}
  \label{eq:87}
   \mathrm  {Herm}({\mathcal H})\ni A\mapsto
     \left  \langle   \rho_{\psi} \Big | A \right  \rangle_{\mathfrak{u}^{*}}=E_{A}(\rho_{\psi}). 
\end{equation}
The pullback of these functions by $\mu$ and $\mu_{{\mathcal P}}$
correspond to the  funtions on ${\mathcal H}$ (resp. ${\mathcal P}{\mathcal H}$)
\begin{equation}
  \label{eq:105}
  \mu^{*}(F_{A}(P_{\psi}))=f_{A}(\psi)=\frac 12 \langle \psi | A \psi\rangle
\end{equation}
and
\begin{equation}
  \label{eq:106}
 \mu_{\mathcal P}^{*}(E_{A}(\rho_{\psi}))=e_{A}(\psi)=\frac {\langle \psi | A\psi\rangle}{2\langle\psi|\psi\rangle} .
\end{equation}

These simple relations allow us to verify that the tensors $G$ and
$\Omega$ on ${\mathcal H}$ (respectively $G_{{\mathcal P}}$ and $\Omega_{{\mathcal
    P}}$ for the  projective case) are $\mu$ (respectively $\mu_{{\mathcal
    P}}$) related to tensors $R$ and $\Lambda$ defined on
$\mathfrak{u}^{*}({\mathcal H})$.  Indeed, if we compute the action of the
tensors $G$ or $\Omega$ on two pullback functions
$f_{A}(\psi)=\mu^{*}(F_{A}(P_{\psi}))$ and
$f_{B}(\psi)=\mu^{*}(F_{B}(P_{\psi})$ for two arbitrary Hermitian
operators $A, B\in \mathrm{Herm}({\mathcal H})$ we obtain:
\begin{equation}
  \label{eq:107}
  G(df_{A},df_{B})(\psi)=f_{A\circ B}(\psi)=\mu^{*}(F_{A\circ
    B}(P_{\psi}))=\mu^{*}\left ( \langle P_{\psi}| A\circ B \rangle_{\mathfrak{u}^{*}}
  \right )=\mu^{*}\left ( R(d\hat A, d\hat B) (P_{\psi})  \right )
\end{equation}
Analagously
\begin{equation}
  \label{eq:107b}
  \Omega(df_{A},df_{B})(\psi)=f_{[A, B]}(\psi)=\mu^{*}(F_{[A,
    B]}(P_{\psi}))=\mu^{*}\left ( \langle P_{\psi} | [A,    B] \rangle_{\mathfrak{u}^{*}} 
  \right )=\mu^{*}\left ( \Lambda(d\hat A, d\hat B) (P_{\psi})  \right ).
\end{equation}

In an analogous way we can see how tensors $R$ and $\Lambda$ are
$\mu_{{\mathcal P}}$--related to tensors $G_{{\mathcal P}}$ and $\Omega_{{\mathcal P}}$:

\begin{multline}
  \label{eq:92}
   \mu_{{\mathcal P}}^{*} \left (\Lambda(d\hat A, d\hat B)(\rho_{\psi})\right )=\mu_{{\mathcal P}}^{*} \left ( \left \langle
  \frac{|\psi\rangle\langle\psi|}{\langle\psi|\psi\rangle} | [A,
  B]\right \rangle_{\mathfrak{u}^{*}} \right ) \\=\mu_{{\mathcal P}}^{*} \left (\frac 12\mathrm{Tr} \left (
   \frac{|\psi\rangle\langle\psi|}{\langle\psi|\psi\rangle}[A, B]
 \right ) \right )=\mu_{{\mathcal P}}^{*} \left (E_{[A,
    B]}(\rho_{\psi})\right )= e_{[A,B]}(\psi)=\Omega_{{\mathcal P}}(\psi)(de_{A}, de_{B}).
\end{multline}

The symmetric tensor is slightly different because of the projective
nature. In any case, it is also straightforward that
\begin{equation}
  \label{eq:93}
  \mu_{{\mathcal P}}^{*}\left (R(d\hat A, d\hat B)(\rho_{\psi})\right ) =\mu_{{\mathcal P}}^{*}(E_{[A\circ
   B]}(\rho_{\psi}))=e_{A\circ B}(\psi)=G_{{\mathcal P}}(de_{A},
 de_{B})(\psi)+e_{A}(\psi)e_{B}(\psi). 
\end{equation}

These relations also determine the equivalence of Schrödinger's and
Heisenberg's pictures when both are meaningful. Indeed, because of the
correspondences between the operators and the functions, it is immediate to
verify that the Hamiltonian vector field $X_{f_{h}}=-\Omega(df_{H},
\cdot)$ whose integral curves correspond to the solutions of
Schrödinger equation is mapped by $T\mu: T{\mathcal H}\to
T\mathfrak{u}^{*}({\mathcal H})$ onto the Hamiltonian vector field
$X_{\hat H}=-\Lambda(d\hat H, \cdot)$ whose integral curves define the
solutions of Heisenberg equation on $\mathfrak{u}^{*}({\mathcal H})$:
\begin{equation}
  \label{eq:94}
  T\mu(X_{f_{H}})=X_{\hat H}.
\end{equation}
Analogously, if we consider the momentum mapping for ${\mathcal P}$,
$T\mu_{{\mathcal P}}:T{\mathcal P}\to T\mathfrak{u}^{*}({\mathcal H})$, we obtain
that the image of the solutions of the projective Schrödinger equation
define also the integral curves of Heisenberg equation:
\begin{equation}
  \label{eq:95}
  T\mu_{{\mathcal P}}({\mathcal X}_{e_{H}})=X_{\hat H}.
\end{equation}
Notice that each momentum map has a different image.
We can proceed analogously with any Hamiltonian vector field on both
spaces. We can also map the gradient vector fields.

If we consider
again our example ${\mathcal H}=\mathbb{C}^{2}$, we identify:
\begin{itemize}
\item the image of the Hamiltonian vector fields:
  \begin{equation}
    \label{eq:96}
    T\mu(X_{f_{k}})=X_{\hat \sigma_{k}}; \qquad k=1, 2,3,
  \end{equation}
\item the image of the gradient vector fields
  \begin{equation}
    \label{eq:97}
    T\mu(Y_{f_{k}})=Y_{\hat \sigma_{k}}:=R(d\hat \sigma_{k}, \cdot );
    \qquad k=1,2,3,
  \end{equation}
\item the image of the vector field $\Delta$, which is the gradient
  vector field associated to the norm:
  \begin{equation}
    \label{eq:98}
    T\mu(\Delta)=Y_{\hat \sigma_{0}}=R(d\hat \sigma_{0}, \cdot),
  \end{equation}
and which is proportional (but not equal) to the natural dilation
vector field ($\Delta_{\mathfrak{u}^{*}}$) of the linear structure of $\mathfrak{u}^{*}({\mathcal
  H})$. The precise relation is
\begin{equation}
  \label{eq:110}
  T\mu(\Delta_{{\mathcal H}})=2\Delta_{\mathfrak{u}^{*}}.
\end{equation}
\end{itemize}

 The image of the vector field $\Gamma$ which is the  Hamiltonian
 vector field corresponding to the norm vanishes because of the
 skew-symmetry of the tensor:
  \begin{equation}
    \label{eq:99}
    T\mu(\Gamma)=X_{\hat \sigma_{0}}=\Lambda(d\hat \sigma_{0}, \cdot)=0.
  \end{equation}

If we consider the vector fields associated to the Hamiltonian
function in both spaces we see immediately how the vector field
associated to the Schrödinger equation on ${\mathcal H}$ (i.e., $X_{f_H}$)
is mapped onto the vector field associated to von Neumann equation on
$\mathfrak{u}^{*}({\mathcal H})$, i.e. $X_{\hat H}$.

The set of vector fields  $\{ X_{\hat \sigma_{\mu}}, Y_{\hat
  \sigma_{\mu}}\}$ with $\mu=0,1,2,3$ do not generate now the Lie algebra
$\mathfrak{gl}({\mathcal H}, \mathbb{C})$, since the Hamiltonian vector field
associated to the  identity operator is identically zero.  
This implies that on the image $\mu({\mathcal H})\subset
\mathfrak{u}^{*}({\mathcal H})$ we can recover the action of the group
$GL({\mathcal H}, \mathbb{C})$ which we found at the level of the
Hilbert space, excepting the global phase change of the
determinant. Notice that if we  consider only the Hamiltonian vector 
fields $\{ X_{\hat \sigma_{\mu}}\}$ with $\mu=0,1,2,3$ they generate the
Lie algebra of the special unitary group $SU({\mathcal H})$,  they
obviously define an
involutive distribution. The corresponding foliation defines the set
of orbits of the coadjoint action of the unitary group on the dual of
its Lie algebra.  We can parametrize the points in
$\mathfrak{u}^{*}({\mathcal H})$ by coordinates $\{ y^{0}, y^{1}, y^{2},
y^{3}\}$ corresponding to
\begin{equation}
  \label{eq:108}
 \mathfrak{u}^{*}({\mathcal H})\ni \rho=\sum_{k}y^{k}\sigma_{k}; \qquad  y^{k}(\rho)=\langle \sigma_{k}| \rho
  \rangle_{\mathfrak{u}^{*}}=\frac 12 \mathrm{Tr}(\sigma_{k}\rho).
\end{equation}
The leaves of the foliation will be two dimensional spheres.

If we ask the trace of the state $\rho\in \mathfrak{u}^{*}({\mathcal H})$
to be equal to one, $y^{0}=\frac 12$.  The set of physical states
correspond then to the points which are contained in the three
dimensional ball
\begin{equation}
  \label{eq:109}
  {\mathcal D}({\mathcal H})=\left \{ \rho\in \mathfrak{u}^{*}({\mathcal H}) |\,\,
    y^{0}=\frac 12; \quad (y^{1})^{2}+(y^{2})^{2}+(y^{3})^{2}\leq
    \frac 34\right \}
\end{equation}

The distribution generated by $\{ X_{\hat \sigma_{\mu}}\}$ is tangent
to any sphere contained in ${\mathcal D}({\mathcal H})$. 
Instead, if we consider the distribution generated by
the gradient vector fields, we can verify immediately that it is not
involutive, since the commutator of two gradient vector fields is a
Hamiltonian vector field and therefore the operation is not
inner. Besides, the distribution generated by the gradient vector
fields is tangent to the surface of the outmost sphere  in ${\mathcal D}({\mathcal H})$, but not to
the spheres in the interior of the ball. With respect to them, in
general it will be transversal.

The situation is different if we consider the projection  $\mu_{{\mathcal
    P}}$. Indeed, it is immediate to notice that vector fields
$\Delta$ and $\Gamma$ are annihilated by the projection $\pi:{\mathcal
  H}\to {\mathcal P}$ and hence on the immage by $\mu_{{\mathcal P}}$ there
are only six generators $\{ X_{\hat \sigma_{k}}, Y_{\hat
  \sigma_{k}}\}$  for $k=1,2,3$ and  where 
$$
T\mu_{{\mathcal P}}({\mathcal X}_{e_{k}})=X_{\hat \sigma_{k}}, \qquad
T\mu_{{\mathcal P}}({\mathcal Y}_{e_{k}})=Y_{\hat \sigma_{k}}
$$

Therefore on the submanifold $\mu_{{\mathcal P}}({\mathcal P})\subset
\mathfrak{u}^{*}({\mathcal H})$ we can consider only the action of the
group $SL({\mathcal H}, \mathbb{C})$, whose Lie algebra is generated by
the vector fields  $\{ X_{\hat \sigma_{k}}, Y_{\hat
  \sigma_{k}}\}$  for $k=1,2,3$. Analogously, when considering the
integrable distribution generated by the Hamiltonian vector fields $\{
T\mu_{{\mathcal P}}({\mathcal X}_{e_{k}})\}$ for $k=1,2,3$, we identify the
orbit of the special unitary group $SU(2)$.

This change in the global group to $SL({\mathcal H})$
when considering the complex projective space (both as a manifold or
as its image by $\mu_{{\mathcal P}}$) is also reflecting a quite remarkable
property, namely the nonlocality of the product $\star$ which encodes
the associative product of operators. Indeed, we already explained
that it is via this product how we can build a $C^{*}$--algebra
structure on the space of functions ${\mathcal E}({\mathcal H})$. Thus we can
write 
\begin{equation}
  \label{eq:111}
  e_{AB}:=e_{A}\star e_{B}=e_{A}e_{B}+\frac 12 G_{{\mathcal P}}(de_{A},
  de_{B})+\frac i2 \Omega_{{\mathcal P}}(de_{A}, de_{B}),
\end{equation}
where we must keep in mind that the product is associative and
non-local.  Thus $({\mathcal E}_{\mathbb{C}}({\mathcal H}), \star)$ (i.e., the
complexification of the algebra of functions generated by Hermitian
operators) becomes a $C^{*}$-algebra.  Now, if we consider the
automorphisms of this algebra, we will identify the whole group
$GL({\mathcal H})$ acting on ${\mathcal E}_{\mathbb{C}}({\mathcal H})$ as
automorphisms with respect to the $\star$ operation, i.e.,
transformations of the type
\begin{equation}
  \label{eq:112}
\Phi_{T}:  e_{A}\mapsto e_{T}\star e_{A}\star e_{T^{-1}}=e_{TAT^{-1}}, \qquad T\in GL({\mathcal H}).
\end{equation}
As this product is non-local, transformations on functions do not
induce transformations on the space on which they are defined. In
other terms, infinitesimal generators of one parameter subgroups will
not be derivations of the pointwise product of functions, therefore we
will not have vector fields associated to them as infinitesimal generators.

\subsubsection{The GNS construction}
We saw in the previous section how the momentum mapping allows us to
map the Schrödinger picture on the Heisenberg one. Let us see now how
the GNS construction allows us to move in the other direction, i.e.,
from the Heisenberg picture we will recover the Schrödinger one.

The starting point is thus a $C^{*}$--algebra ${\mathcal A}$ which
contains the set of physical observables as a real subspace. States
are represented as  normalized positive linear functionals $\omega$
which satisfy 
\begin{equation}
  \label{eq:1}
  \omega(a^{*}a)\geq 0 ;\qquad  \omega(\mathbb{I})=1.
\end{equation}
Therefore, we can embed the set of states ${\mathcal D}(\mathcal A)$ in the
dual of ${\mathcal A}$.

Each state $\omega$ allows us to introduce a pairing between the
elements of ${\mathcal A}$:
\begin{equation}
  \label{eq:2}
  \langle a, b\rangle_{\omega}=\omega(a^{*}b).
\end{equation}
The pairing is positive  because of the properties of $\omega$
but it may be degenerate.  We define then the Gelfand ideal ${\mathcal
  I}_{\omega}$ to be the kernel:
\begin{equation}
  \label{eq:3}
  {\mathcal I}_{\omega}=\{ a\in {\mathcal A}\, | \,\,\omega(a^{*}a)=0 \}
\end{equation}
If we define the quotient space $\tilde {\mathcal H}_{\omega}={\mathcal A}/{\mathcal I_{\omega}}$  and the subsequent
equivalence classes 
\begin{equation}
  \label{eq:4}
  \Psi_{a}=\{a+ \alpha\,|\,\,a\in {\mathcal A};  \alpha\in {\mathcal I}_{\omega}\},
\end{equation}
we can define a structure of pre-Hilbert space on $\tilde {\mathcal H}_{\omega}$
by using the scalar product
\begin{equation}
  \label{eq:11}
  \langle \Psi_{a}|\Psi_{b}\rangle=\omega(a^{*}b).
\end{equation}
By completing $\tilde {\mathcal H}_{\omega}$ with respect to the
corresponding norm topology, we define a Hilbert space ${\mathcal
  H}_{\omega}$. 

On this Hilbert space we can define a representation of the
$C^{*}$--algebra ${\mathcal A}$ in the form:
\begin{equation}
  \label{eq:30}
  \pi_{\omega}:{\mathcal A}\times {\mathcal H}_{\omega}\to {\mathcal H}_{\omega};
  \qquad \pi_{\omega}(b)\Psi_{a}=\Psi_{ab}. 
\end{equation}
In more formal terms we have
\begin{definition}
  A *-representation of $\mathcal{A}$ on the Hilbert space
$\mathcal{H}_{\omega}$ is a homomorphism $\pi_{\omega}$ from $\mathcal{A}$ to the algebra
$\mathcal{B}(\mathcal{H_{\omega}})$  of 
bounded operators on $\mathcal{H}_{\omega}$ which maps the involution of 
$\mathcal{A}$ on the adjoint operation of $\mathcal{B}(\mathcal{H}_{\omega})$ .
\end{definition}

If we consider the vector associated to the identity element of ${\mathcal
  A}$, i.e.  
$$
|\Omega\rangle=\Psi_{\mathbb{I}},
$$
we can recover the state $\omega$ as 
\begin{equation}
  \label{eq:31}
  \omega(a)=\langle \Omega| \pi_{\omega}(a)\Omega\rangle, \qquad
  \forall a\in {\mathcal A}.
\end{equation}

As we know that the set of states ${\mathcal D}({\mathcal A})$ can be embedded
in ${\mathcal A}$, the expression above implies that the Hilbert space
${\mathcal H}_{\omega}$ may be thought of as a subspace of ${\mathcal A}$
coinciding with the orbit of the left action of ${\mathcal A}$ on itself  which passes
through $\omega$, once identified with an element of ${\mathcal A}$
(remember that we are in finite dimension):
\begin{equation}
  \label{eq:32}
  {\mathcal H}_{\omega}\simeq {\mathcal O}_{{\mathcal A}}(\omega).
\end{equation}

This can also be done for any other element of ${\mathcal H}_{\omega}$:
\begin{definition}
Given $\pi:\mathcal{A}\to \mathcal{B}(
\mathcal{H_{\omega}})$, a vector $\xi\in \mathcal{H}_{\omega}$ is said to be \textbf{cyclic}
if the orbit
$$
\mathcal{O}_\xi=\{ \pi_{\omega}(x) \xi | x\in \mathcal{A}\}
$$ 
is dense in $\mathcal{H}_{\omega}$.  If a cyclic vector exists, we say that
$\pi_{\omega}$ is a \textbf{cyclic representation}. If we consider a normalized
vector  $\|\xi\|=1$, the functional
$$
\rho_{\xi}: x\mapsto \langle \xi |\pi_{\alpha}(x) \xi\rangle\in \mathbb{C}
$$
is a state of  $\mathcal{A}$.
  
\end{definition}
In this context we notice that the ambiguity in the Hermitian structure of
the corresponding Hilbert space is connected with the ambiguity in the
choice of the starting fiducial state to define the GNS Hilbert space and
the subsequent realization of the unitary group. The abstract group is
always the same (the unitary group) but the realizations may be different.

Besides, from the state $\omega$ (i.e., from $|\Omega)$) we can define
other states  as density matrices $\rho$  defined on ${\mathcal B}({\mathcal H}_{\omega})$
  as the convex combinations of projectors on one dimensional
  subspaces of ${\mathcal H}_{\omega}$:
  \begin{equation}
    \label{eq:39}
    \omega_{\rho}(a)=\mathrm{Tr}(\rho \pi_{\omega}(a)).
  \end{equation}
These generalized states lead to representations of ${\mathcal A}$ which
are reducible and decomposable as a direct sum:
\begin{equation}
  \label{eq:33}
  \pi_{\rho}=\bigoplus_{\alpha}\pi_{\alpha}
\end{equation}
on subspaces
\begin{equation}
  \label{eq:70}
  {\mathcal H}_{\rho}=\bigoplus_{\alpha}{\mathcal H}_{\alpha}
\end{equation}
The vacuum state $|\Omega_{\rho}\rangle$ corresponding to the identity
element of ${\mathcal A}$ decomposes then as a sum:
\begin{equation}
  \label{eq:71}
  |\Omega_{\rho}\rangle=\sum_{\alpha}|\Omega_{\alpha}\rangle; \qquad
  |\Omega_{\alpha}\rangle\in {\mathcal H}_{\alpha},
\end{equation}
in such a way that the irreducible representations $\pi_{\alpha}$ are
associated to pure states $\xi_{\alpha}$:
\begin{equation}
  \label{eq:73}
\xi_{\alpha}(a)=\frac 1{p_{\alpha}} \langle \Omega_{\alpha}|\pi_{\alpha} |\Omega_{\alpha}\rangle,
\end{equation}
where $p_{\alpha}=\langle\Omega_{\alpha}|\Omega_{\alpha}\rangle$. As
$|\Omega_{\rho}\rangle$ is normalized, $\sum_{\alpha}p_{a}=1$. But
then we can write the state $\rho$ as a convex
combination
\begin{equation}
  \label{eq:74}
  \rho=\sum_{\alpha}p_{\alpha}\xi_{\alpha},
\end{equation}
where $\{\xi_{\alpha}\}$ are pure states.

Notice that, from a geometric point of view,  once we have fixed a
state $\omega$, we can reproduce the analysis of the momentum map
which we considered in the previous section at the level of the Hilbert space
${\mathcal H}_{\omega}$.  Indeed, we can consider the unitary group
$U({\mathcal H}_{\omega})$ and its defining action on  ${\mathcal H}_{\omega}$:
\begin{equation}
  \label{eq:40}
  \Phi:U({\mathcal H}_{\omega})\times {\mathcal H}_{\omega}\to {\mathcal
    H}_{\omega}; \qquad \Psi_{a}\mapsto U\Psi_{a}, 
\end{equation}
and on the corresponding projective space 
\begin{equation}
  \label{eq:40b}
  \Phi_{{\mathcal P}}:U({\mathcal H}_{\omega})\times {\mathcal PH}_{\omega}\to
  {\mathcal PH}_{\omega}; \qquad [\Psi_{a}]\mapsto [U\Psi_{a}].
\end{equation}

Associated to these actions we can define two projections in 
analogy to what we did in the previous section:
\begin{equation}
  \label{eq:69}
  \mu_{\omega}:{\mathcal H}_{\omega}\to \mathfrak{u}^{*}({\mathcal
    H}_{\omega}); \qquad \mu_{\omega}(\Psi_{a})=\frac 12|\Psi_{a}\rangle\langle\Psi_{a}|,
\end{equation}
and
\begin{equation}
  \label{eq:69b}
  \mu_{\omega}^{{\mathcal P}}:{\mathcal PH}_{\omega}\to \mathfrak{u}^{*}({\mathcal
    H}_{\omega}); \qquad \mu^{{\mathcal
      P}}_{\omega}([\Psi_{a}])=\frac{|\Psi_{a}\rangle\langle\Psi_{a}|}{2\langle
    \Psi_{a}|\Psi_{a}\rangle}.
\end{equation}

$\mu_{\omega}$ defines (as in the case of the previous section) a
symplectic realization on the symplectic vector space ${\mathcal
  H}_{\omega}$ of the Poisson manifold $\mathfrak{u}^{*}({\mathcal
    H}_{\omega})$. As ${\mathcal H}_{\omega}$  is a vector space, it is
  called a Hilbert space realization.  $\mu_{\omega}^{{\mathcal
      P}}$ also defines a symplectic realization, but in this case we
  call it  a Kählerian realization.

\subsubsection{Further comments}
Within the study of the $C^{*}$--algebraic approach we can identify
transformation groups given by: 

\begin{itemize}
\item automorphisms of the Jordan structure;

\item automorphisms of the Lie structure;
\item automorphisms of the complex structure.
\end{itemize}
The action of the unitary group  preserves all of them.

From what we have said up to now,  it should be clear the $C^{*}$--
algebraic formalism appears to be 
more general than the Schrödinger formalism which emerges via the GNS
construction when we use a pure state as a starting fiducial state.

In our running example, the $C^*$ algebra is the  complex general algebra
of matrices $M_{n}(\mathbb{C})\sim \mathfrak{gl}(n, \mathbb{C})$. When
we consider it as a Lie algebra and exponentiate it we get 
the complex general Lie group $GL(n, \mathbb{C})$. This group contains
as a maximal compact 
subgroup the unitary group $U(n)$,  whose complexification gives back the general
Lie group. When the group is realized as a group of matrices,
similarity transformations with respect to the linear group will take from one realization of $U(n)$
to a different realization, and consequently, from one Hermitian
structure to a different one on the vector space carrying the given realization.
 In finite dimensions the complex linear group and the unitary group  are in
one-to-one correspondence and one determines the other. Of course in infinite
dimensions the situation is far from being so simple and things are not
completely well defined due to the different topologies available and the
lack of a properly defined differential geometry along with a proper
definition of infinite dimensional Lie group. In any case, in finite
dimensions the two groups do determine each other. As we have argued, for the
complex projective space (the space of pure states), i.e., the two
dimensional sphere for the two-level 
system, both groups act transitively on it. The unitary group acts by
preserving the relevant structures while the complexification does not.

Within the space of  mixed states the two groups have different
orbits. The stratification by the rank which is natural on the space of
mixed states shows that each stratum is the union of different orbits of the
unitary group while each stratum is an orbit of a proper defined action of
the general complex linear group which acts in a non linear manner to
preserve the trace.

The infinitesimal generators of this action have an important interpretation
in the framework of the Markovian dynamics for open quantum systems. 
Therefore we maintain that, in the realistic setting of open quantum systems,
the complexification of the unitary group seems to be more relevant than
the unitary group itself. With this claim in mind, it is quite natural to
consider this group as the relevant group of quantum mechanics in the
ideology of the Klein's programme.

In addition, if we priviledge the $C^{*}$--algebra approach, the
general complex linear group emerges also as the exponentiated version
of the $C^{*}$--algebra thought of as a Lie algebra. In the following section we shall
investigate the geometrical aspects of the general linear group.

\section{$\mathrm{GL}(\mathcal{H})$ and Kraus operators}

\subsection{Geometry of $\mathrm{GL}(\mathcal{H})$}
As we just saw the group $\mathrm{GL}(\mathcal{H})$ seems to  arise naturally
from the geometric structures of an open quantum system. Let us study its
geometry in more detail. As in the geometrical approach we are
considering the Hilbert space as a real vector space, our ingredients
are a vector space ${\mathcal H}$,  a linear space structure encoded in a
Liouville operator $\Delta$ and a complex structure $J$ ( i.e., a
$(1,1)$ tensor field $J$ which satisfies that $J^{2}=-\mathbb{I}$) 
compatible with $\Delta$ in the sense that 
\begin{equation}  \label{eq:5}
\mathcal{L}_{\Delta}J=0.
\end{equation}

We can consider the group $\mathrm{GL}(\mathcal{H})$ as the (finite
dimensional) subgroup of the diffeomorphism group of the set $\mathcal{H}$
which keeps invariant the Liouville operator $\Delta$ and the complex
structure  $J$. Thus 
\begin{equation}  \label{eq:6}
\mathrm{GL}({\mathcal H})=\left \{ \phi\in \mathrm{Diff}({\mathcal H}) \vdash
\phi_{*}\Delta=\Delta; \phi_{*}J=J \right \}
\end{equation}
As a consequence, these transformations preserve also $J(\Delta)= \Gamma$.
The information on the projective space is already encoded here
since it arises as the foliation generated by the integrable distribution
generated by $\Delta$ and $J(\Delta)\coloneqq \Gamma$ (the global phase
generator). 

\subsubsection{Properties of the Lie groups}
The following facts are easy to prove
\begin{itemize}
\item $ \mathrm{GL}(\mathcal{H})$ contains
as a maximal compact  subgroup the unitary group $U(n)$,
\item the complexification of the group $\mathrm{U}(\mathcal{H})$ is isomorphic to $%
\mathrm{GL}(\mathcal{H})$.

\item In addition, introducing the tangent and cotangent bundles of
  the unitary group, we have
\begin{equation}  \label{eq:7}
\mathrm{GL}(\mathcal{H})\rightleftarrows T\mathrm{U}(\mathcal{H}) \rightleftarrows T^{*}\mathrm{U%
}(\mathcal{H})
\end{equation}
All three groups  are symplectomorphic, the structure on $GL({\mathcal
  H})$ being defined from     the product of $U({\mathcal
  H})$ and the Borel subgroup $ B(n, \mathbb{C})$ (see \cite{Alekseevsky1998}).

\end{itemize}
We shall denote as $%
\mathfrak{gl}(\mathcal{H})$ and $\mathfrak{u}(\mathcal{H})$ the corresponding Lie
algebras of these Lie groups.

\subsubsection{Properties of the Lie algebras}
In what regards the Lie algebra $\mathfrak{gl}(\mathcal{H})$ we have also some
interesting properties which will be useful later.

\begin{itemize}
\item As a matrix algebra, $\mathfrak{gl}(\mathcal{H})$ is an associative
algebra with involution, corresponding to the adjoint operation $A\mapsto
A^{\dagger}$.

\item Second,  $\mathfrak{gl}(\mathcal{H})$ carries  a Hilbert space
structure defined by the scalar product 
\begin{equation}  \label{eq:8}
\langle A |B\rangle=\mathrm{Tr}(A^{\dagger}B)
\end{equation}
If we restrict this operation to the subalgebra of the unitary group
$\mathfrak{u}({\mathcal H})$, it is immediate that it defines a
non-degenerate metric. Therefore it can be used to define an isomorphism between the
algebra and its dual
\begin{equation}
  \label{eq:10}
 \hat{}:  \mathfrak{u}({\mathcal H})\to \mathfrak{u}^{*}({\mathcal H}); \qquad
 A\mapsto \hat A:=\langle A, \cdot \rangle \in \mathfrak{u}^{*}({\mathcal H}).
\end{equation}
This is the inverse mapping with respect to Eq. (\ref{eq:83}).
\item the diffeomorphisms connecting the group
  $\mathrm{GL}(\mathcal{H})$ and the tangent and 
cotangent bundle of the unitary group $\mathrm{U}(\mathcal{H})$ have a simple
translation at the level of Lie algebras: 
\begin{equation}  \label{eq:9}
\mathfrak{gl}(\mathcal{H})\rightleftarrows T\mathfrak{u}(\mathcal{H}) \rightleftarrows T%
\mathfrak{u}^{*}(\mathcal{H})\rightleftarrows T^{*}\mathfrak{u}^{*}(\mathcal{H})
\rightleftarrows T^{*}\mathfrak{u}(\mathcal{H}).
\end{equation}
The new equivalences arise from the isomorphism of the unitary algebra and
its dual coming from the invertibility of the metric structure defined
by Eq. (\ref{eq:10}).
\end{itemize}
\subsection{The space of density states $\mathcal{D}(\mathcal{H})$}

Let us consider now the structure and properties of the space of physical
states $\mathcal{D}(\mathcal{H})$. We know that such a space is defined as a
subset of the dual of the unitary Lie algebra $\mathfrak{u}(\mathcal{H})$,
defined as those matrices in $\mathfrak{u}^{*}(\mathcal{H})$ which are
positive definite and have trace equal to one, i.e., 
\begin{equation}  \label{eq:35}
\mathcal{D} (\mathcal{H})=\left \{ \rho\in \mathfrak{u}^{*}(\mathcal{H})
|\,\,\, \rho>0 ; \mathrm{Tr} \rho=1 \right\}.
\end{equation}

It is convenient to define first the set of positive matrices $\mathfrak{P}(%
\mathcal{H})$ and impose then the constraint on the trace. Thus, we consider 
\begin{equation}  \label{eq:36}
\mathfrak{P}(\mathcal{H})=\left \{ \omega \in \mathfrak{\ u}^{*}(\mathcal{H}) |
\,\, \omega =RR^{\dagger} \quad R\in \mathfrak{gl}(\mathcal{H})\right \}.
\end{equation}
This view allows to derive many properties in simple terms. For instance,
the left action of the group $GL(\mathcal{H})$ 
is then easily written: 
\begin{equation}  \label{eq:37}
\phi: GL(\mathcal{H})\times \mathfrak{P}(\mathcal{H})\to \mathfrak{P}(%
\mathcal{H}); \qquad (g, \omega)\mapsto g\omega g^{\dagger}.
\end{equation}
On the other hand, the right action of the unitary subgroup projects
onto the identity, showing that $\mathfrak{P}(H)$ is the base manifold
of a $U({\mathcal H})$--bundle.

When considered as an action on $\mathfrak{u}^{*}(\mathcal{H})$, $\phi$
changes the spectrum of $\pi$, but it preserves its rank (signature).
Indeed, it is simple to see that, along the orbit, the number of positive,
negative and null eigenvalues of the elements is preserved. When considered
as a bi-linear form, this property corresponds to Sylvester's Law of inertia
generalized to complex vector spaces (see \cite{Sylvester4} for the original
proof and also \cite{Grabowski2005a}). If we consider the action on the set
of positive operators, it is obvious that the action is inner. We can
consider thus a decomposition of $\mathfrak{P}({\mathcal H})$ according to
the rank of the state  and define
\begin{equation}
  \label{eq:28}
  \mathfrak{P}^{k}({\mathcal H})=\left \{ \omega\in \mathfrak{u}^{*}({\mathcal
      H}) | \, \omega=RR^{\dagger}, \,\, R\in \mathfrak{gl}({\mathcal H});\,\,
    \mathrm{rank}(\omega)=k \right \}.
\end{equation}
Therefore $\mathfrak{P}^{k}$ contains those positive elements in
$\mathfrak{u}^{*}({\mathcal H})$ having $k$ positive eigenvalues, the rest
being zero. On it, we can consider also the condition on the trace and
define the corresponding subset of the set ${\mathcal D}({\mathcal H})$:
\begin{equation}
  \label{eq:28b}
  \mathcal{D}^{k}({\mathcal H})=\left \{ \rho\in\mathfrak{P}^{k} ({\mathcal
      H}) | \,  \,\,\mathrm{Tr}(\rho)=1\right \}.
\end{equation}

The problem becomes more complicated  when we want to consider the action of the group on
the set of density operators $\mathcal{D}(\mathcal{H})$. It is simple to see
that the action (\ref{eq:37}) does not preserve $\mathcal{\ D}(\mathcal{H})$%
. Indeed, for a general element $g\in GL(\mathcal{H})$ 
\begin{equation}  \label{eq:34}
\mathrm{Tr} \left (g\rho g^{\dagger}\right )\neq \mathrm{Tr} \rho.
\end{equation}
Therefore, we must modify the action $\phi$ to define an inner operation on
the set of density states: 
\begin{equation}  \label{eq:38}
\phi^{\mathcal{D}}:GL(\mathcal{H})\times \mathcal{D}(\mathcal{H})\to 
\mathcal{D}(\mathcal{H}); \qquad (g, \rho)\mapsto \frac{g \rho g^{\dagger}}{%
\mathrm{Tr}\left (g\rho g^{\dagger}\right )}
\end{equation}
As the action on $\mathfrak{P}(\mathcal{H})$ preserves the rank, we know
that the denominator can not vanish and therefore $\phi^{\mathcal{D}}$ is
well defined. It is immediate to verify that the action preserves the
sets ${\mathcal D}^{k}$, i.e
\begin{equation}
  \label{eq:79}
  \phi^{{\mathcal D}}:GL(\mathcal{H})\times \mathcal{D}^{k}(\mathcal{H})\to 
\mathcal{D}^{k}(\mathcal{H}),
\end{equation}
since the action $\phi$ preserves, by definition, the subsets of positive
operators $\mathfrak{P}^{k}({\mathcal H})$.  Both manifolds ${\mathcal
  D}({\mathcal H})$ and $\mathfrak{P}({\mathcal H})$ become thus
\textbf{stratified manifolds}, the strata being the subsets ${\mathcal
  D}^{k}({\mathcal H})$ and $\mathfrak{P}^{k}({\mathcal H})$, i.e., the orbits
of the corresponding action of the general linear group $GL(n,
\mathbb{C})$.

The set ${\mathcal D}^{1}({\mathcal H})$ corresponds to those Hermitian operators which
have only one eigenvalue different from zero, and therefore equal to
one. These are clearly the projectors on one-dimensional subspaces of
the Hilbert space ${\mathcal H}$, and hence they are in one-to-one
correspondence with the points of the projective space ${\mathcal P}$ by
the momentum map $\mu_{{\mathcal P}}$ introduced in Eq. (\ref{eq:85}). We
studied in previous Section how the Hamiltonian and gradient vector
fields defined different types of transformations on ${\mathcal
  D}^{1}({\mathcal H})$. 
As we saw, the set
of Hamiltonian vector fields  were the
fundamental vector fields corresponding to the action of the special unitary
group (subgroup of $GL({\mathcal H}, \mathbb{C})$) while together with the
gradient vector fields they were the
fundamental vector fields of the action of the special linear group
$SL({\mathcal H}, \mathbb{C})$. It is not possible to consider the action
of the other generators of $GL({\mathcal H}, \mathbb{C})$ on ${\mathcal
  D}^{1}({\mathcal H})$, since the action of the generators associated to
$\hat \sigma_{0}$ (the vector fields which are associated via the
momentum map $\mu$ with vector fields $\Delta$ and $\Gamma$) act
trivially on ${\mathcal D}^{1}({\mathcal H})$.

A final comment is necessary with respect to the complex structure.
On the projective space,  or equivalently on the subset ${\mathcal
  D}^{1}({\mathcal H})$, there is a canonical complex
structure. Therefore, we know that gradient and Hamiltonian vector
fields are in one-to-one correspondence defined by the tensor $J$.
Therefore, the whole $SL({\mathcal H}, \mathbb{C})$--orbit ${\mathcal
  D}^{1}({\mathcal H})$  can be obtained from the complexification of the
generators of the unitary group (the Hamiltonian vector fields). The
situation for other strata is more complicated since, although we can
still obtain the stratum from the Hamiltonian and gradient vector
fields,  we lack of a properly defined complex structure on them to
relate both sets (see \cite%
{Grabowski2005a, Grabowski2006}). 

\subsection{The set of density states $\mathcal{D}(\mathcal{H})$ and Kraus operators}

As we saw in the previous section, 
the left action of $\mathfrak{gl}(\mathcal{H})$ on itself corresponding to the
associative product 
\begin{equation}  \label{eq:12}
\cdot: \mathfrak{gl}(\mathcal{H})\times \mathfrak{gl}(\mathcal{H})\to \mathfrak{gl}(\mathcal{H}); \qquad
(A_{1}, A_{2})\mapsto A_{1}A_{2},
\end{equation}
can be projected on $\mathcal{P}(\mathcal{H})$: 
\begin{equation}  \label{eq:13}
\cdot:\mathfrak{gl}(\mathcal{H})\times \mathcal{P}(\mathcal{H})\mapsto \mathcal{P}(\mathcal{H}); \qquad
(M, \rho)\mapsto M\rho M^{\dagger}
\end{equation}

Given a family of operators $\mathbb{M}=\{ M_{1}, \cdots, M_{m}\}\in 
\mathfrak{gl}(\mathcal{H})\times \cdots \times \mathfrak{gl}(\mathcal{H})$, we can consider the
action on density states as 
\begin{equation}  \label{eq:14}
( \mathbb{M}, \rho)\mapsto \sum_{j}M_{j}\rho M_{j}^{\dagger}\coloneqq{\mathcal K}%
_{\mathbb{M}}(\rho).
\end{equation}

The operator $\mathcal{K}_{\mathbb{M}}$ shall be called a \textbf{Kraus
map}. The Kraus map is said to be normalized if 
\begin{equation}  \label{eq:18}
\sum_{k}M_{k}^{\dagger }M_{k}=\mathbb{I}.
\end{equation}

\begin{proposition}
  The composition of two Kraus maps is an inner operation.  The
  set of Kraus maps endowed with the composition operation
  becomes a semi-group.
  \begin{equation}
    \label{eq:15}
    {\mathcal K}_{\mathbb M}\circ {\mathcal K}_{\mathbb M'}
    =\sum_{jk}(M_{j}M'_{k})\rho(M_{j}M'_{k})^{\dagger}={\mathcal
      K}_{\mathbb {M M'}}.
  \end{equation}
\end{proposition}

By using the Hilbert space structure of $\mathfrak{gl}(\mathcal{H})$, we can realize a
Kraus map as a sum (Jamiolkowski isomorphism \cite{Jamiolkowski1972275}%
): 
\begin{equation}  \label{eq:16}
\mathcal{K}_{\mathbb{M}}=\sum_{j}|M_{j}\rangle\langle M_{j}|.
\end{equation}

At this point we can claim 
\begin{proposition}
  The operators of $\GL(\mathcal{H})$ form the largest subgroup of the semi-group
  of normalized Kraus maps.
\end{proposition}
\begin{proposition}
  If a Kraus map ${\mathcal K}_{\mathbb M}$ is invertible inside the
  set of Kraus maps, there exists an element $M\in \GL(\mathcal{H})$ such
  that
  \begin{equation}
    \label{eq:17}
    {\mathcal K}_{\mathbb M}\rho=M \rho M^{\dagger}.
  \end{equation}
\end{proposition}

\subsection{Markovian dynamics} 
In this context we can also consider the dynamics for general
Markovian dynamics. It is well known that the most general markovian
dynamical system takes the form of the Kossakowski-Lindblad
superoperator  $L$ defined as the infinitesimal generator of a
one-parameter semigroup of transformations on $\mathfrak{u}^{*}({\mathcal H})$ (see
\cite{Lindblad1976,Gorini1976b}) which takes the form 
\begin{equation}
  \label{eq:101}
  L:\rho\mapsto L(\rho):= -i[H, \rho]+\frac 12
  \sum_{ij=1}^{N^{2}-1}c_{ij}\left ( [F_{i},\rho
    F_{j}^{\dagger}]+[F_{i}\rho,F_{j}^{\dagger} ]\right ), 
\end{equation}
where $c_{ij}$ defines a complex positive matrix, $H$ is Hermitian and
$F_{k}$ is traceless and satisfies that 
\begin{equation}
  \label{eq:102}
  \mathrm{Tr}(F_{i}F_{j}^{\dagger})=\delta_{ij}.
\end{equation}
It is simple to see that such a transformation preserves the trace of
$\rho$ but changes its spectrum.

We can rewrite the expression of $L$ in more
geometrical terms and exhibit some of its properties in a more
explicit way. Indeed, from Eq. (\ref{eq:101}) it is immediate that the
first factor of the right hand side can be understood as  a
Hamiltonian vector field. The last two terms are more difficult to
interpret in the form we wrote them. But let us consider an
alternative factorization by re-writting them as
$$
[F_{i},\rho   F_{j}^{\dagger}]+[F_{i}\rho,F_{j}^{\dagger} ]=
F_{i}\rho F_{j}^{\dagger}-\rho F_{j}^{\dagger}F_{i}+F_{i}\rho
F_{j}^{\dagger}-F_{j}^{\dagger} F_{i}\rho=-[F_{j}^{\dagger}F_{i},
\rho]_{+}+ 2F_{i}\rho F_{j}^{\dagger}
$$
If we consider a basis where the matrix $c_{ij}$ is equal to the identity, and
denote the corresponding eigenvectors written in terms of the
operators $\{F_{k} \}$ as $V_{\alpha}$, we can write the generator $L$ in
the form
\begin{equation}
  \label{eq:103}
  L(\rho)=-i[H, \rho]-\frac 12 \sum_{\alpha= 1}^{N^{2}-1}\left (
  [V_{\alpha}^{\dagger}V_{\alpha}, \rho]-2V_{\alpha}\rho
  V_{\alpha}^{\dagger} \right )=
-i[H, \rho]-\frac 12 [G, \rho]_{+}+\sum_{\alpha= 1}^{N^{2}-1} V_{\alpha}\rho
  V_{\alpha}^{\dagger},
\end{equation}
where 
\begin{equation}
  \label{eq:104}
  G=\sum_{\alpha}V_{\alpha}^{\dagger}V_{\alpha}.
\end{equation}

We easily recognize in Eq. (\ref{eq:103}) the three types of
transformations we have presented in this paper: 
\begin{itemize}
\item the first term defines a Hamiltonian vector field,
\item the second corresponds to a gradient vector field
\item and finally, the third represents the action of the Kraus
  map $\mathbb{V}=\{ V_{1}, \cdots, V_{N^{2}-1}\}$.
\end{itemize}

The Hamiltonian vector field alone generates a unitary
transformation. The second  and third terms are the ones responsible
for the breaking of unitarity in the markovian evolution. But they
break unitarity in such a way that  the total transformation generated by $L$ does
not change the trace of $\rho$.

Summarizing we can say that the generators of the actions of the group
$\mathrm{GL}({\mathcal H})$ on ${\mathcal D}({\mathcal H})\subset \mathfrak{u}^{*}({\mathcal
  H})$ allow us to express in geometrical terms the most general form
of markovian evolution of a finite dimensional quantum system.

\section{Conclusions and outlook}

We have seen that the markovian dynamical evolution of an open system is not
associated with a group of transformations but with a semi-group. The
maximal subgroup is the general linear group which we interpret as the
relevant group of Quantum Mechanics and the one identified according
to the Klein programme when dealing with open systems. A natural
question arises: is it possible to generalize Klein's programme to
semi-groups?

Another interesting possibility arises from the fact that the $C^{*}$
--algebra which plays a crucial role in our presentation is also a groupoid algebra
(see references \cite{Ibort2013a} ). Is it
possible to consider an extension of the Erlangen 
Programme to groupoids?


\end{document}